\documentclass[a4paper,fleqn,usenatbib]{mnras}

\usepackage{newtxtext,newtxmath}

\usepackage[T1]{fontenc}
\usepackage{ae,aecompl}

\usepackage[dvipdfmx]{graphicx}	
\usepackage{amsmath}	
\usepackage{amssymb}	
\usepackage{indentfirst}	
\usepackage{pict2e}
\usepackage{color}
\usepackage{bm}

\usepackage{ulem}

\title[Condition for dust evacuation from the first galaxies]{ 
Condition for dust evacuation from the first galaxies
}

\author[Fukushima et al.]{
Hajime Fukushima$^{1}$\thanks{E-mail:fukushima@tap.scphys.kyoto-u.ac.jp}
Hidenobu Yajima$^{2,3}$
Kazuyuki Omukai$^{2}$
\\
$^{1}$Department of Physics, Kyoto University, Sakyo, Kyoto 6060-8502, Japan\\
$^{2}$Astronomical Institute, Tohoku University, Aoba, Sendai 980-8578, Japan\\
$^{3}$Frontier Research Institute for Interdisciplinary Sciences, Tohoku University, Sendai, Miyagi 980-8578, Japan
}

\date{Accepted XXX. Received YYY; in original form ZZZ}

\pubyear{2018}

\begin{document}
\label{firstpage}
\pagerange{\pageref{firstpage}--\pageref{lastpage}}
\maketitle

\begin{abstract}
Dust enables low-mass stars to form from low-metallicity gas by inducing fragmentation of clouds via the cooling by its thermal emission. 
Dust may, however, be evacuated from star-forming clouds due to radiation force from massive stars.
We here study the condition for the dust evacuation by comparing the dust evacuation time with the time of cloud destruction due to either expansion of H{\sc ii} regions or supernovae.
The cloud destruction time has weak dependence on the cloud radius, while the dust evacuation time becomes shorter for a cloud with the smaller radius.
The dust evacuation thus occurs in compact star-forming clouds whose column density is $N_{\rm H} \simeq 10^{24} - 10^{26} ~{\rm cm^{-2}}$.
The critical halo mass above which the dust evacuation occurs becomes lower for higher formation redshift, e.g., $\sim 10^{9}~M_{\odot}$ at redshift $z \sim 3$ and $\sim 10^{7}~M_{\odot}$ at $z \sim 9$.
In addition, metallicity of the gas should be less than $\sim 10^{-2} ~ Z_{\odot}$. 
Otherwise the dust attenuation reduces the radiation force significantly.
From the dust-evacuated gas, massive stars are likely to form even with metallicity above $\sim 10^{-5}~Z_{\odot}$, the critical value for low-mass star formation due to the dust cooling.
This can explain the dearth of ultra-metal poor stars with the metallicity lower than $\sim 10^{-4}~Z_{\odot}$. 
\end{abstract}

\begin{keywords}
stars: formation -- stars: Population II -- stars: low-mass -- galaxies: evolution -- dust, extinction 
\end{keywords}


\section{Introduction}\label{introduction}

The initial mass function (IMF) of stars deeply influences subsequent galaxy evolution and star formation history. 
Stellar feedback from massive stars alter the physical state and structure of the interstellar medium \citep[e.g.][]{McKee1977}.
Supernovae (SNe) provide the interstellar medium with heavy elements and foster chemical evolution of galaxies.
On the other hand, low mass stars account for the most of stellar mass of aged galaxies due to their longevity. 
Even primordial low-mass stars, if formed, can still survive as main sequence stars in the Galaxy \citep[e.g.][]{Machida2008,Clark2011}.

The mass of stars likely depends on the metallicity of the forming environments. 
At the solar metallicity, the IMF has the peak around $0.4~M_{\odot}$, which means the most stars are low-mass objects \citep{Kroupa2001,Chabrier2003}.
On the other hand, the IMF of primordial stars was theoretically predicted with recent radiative-hydrodynamics simulations in the cosmological context. \citep{Hirano2014,Hirano2015, Susa2014, Stacy2016, Hosokawa2016}. 
Those numerical simulations suggest that the primordial stars were formed in the mass range in $\sim 10~M_{\odot}$ - a few 100 $M_{\odot}$ with the typical mass being much larger than that of the sun.

This transition in the characteristic stellar mass from massive to low-mass objects can be ascribed to the change in fragmentation mass of star-forming 
clumps with accumulation of metals \citep{Omukai2000, Bromm2001}. 
In literatures, the transitional metallicity is often called the critical metallicity $Z_{\rm crit}$. 
Fragmentation of clumps is induced by the dust cooling \citep{Schneider2002} as well as metal-line cooling \citep{Bromm2001}.
With gas-phase metals, clumps fragment into massive dense cores of no smaller than a few $10~M_{\odot}$ even with $Z \ga 10^{-3.5} ~Z_{\odot}$ \citep{Bromm2003,Santoro2006}.
On the other hand, the fragmentation mass becomes as small as $0.1-1~M_{\odot}$ in the case of taking into account the dust cooling for $Z \ga 10^{-6} - 10^{-5}~Z_{\odot}$ \citep{Omukai2005,Schneider2003} in the case that the dust depletion factor, which is defined by
\begin{eqnarray}
	f_{\rm dep} \equiv \frac{M_{\rm dust}}{M_{\rm dust} + M_{\rm metal}},  \label{1002.1}
\end{eqnarray}
where $M_{\rm metal}$ is the metal mass in the gas phase and $M_{\rm dust}$ is that taken into dust grains, is the same as in the Galaxy.
By converting this metallicity into dust-to-gas ratio $\mathcal D$, the critical dust-to-gas ratio is $\mathcal D_{\rm crit} = \left[2.6-6.3 \right]\times 10^{-9}$ \citep{Schneider2012}.
Therefore, the dust cooling is supposed to produce low-mass stars of $\lesssim 1~M_{\odot}$ even at such low metallicity as $10^{-6} - 10^{-5}~Z_{\odot}$. 

Recent survey observations successfully discovered several metal-poor stars with metallicity lower than $\sim 10^{-4}~Z_{\odot}$, but these ultra-metal poor stars are very rare in the Galaxy \citep[e.g.][]{Frebel2015}. 
\citet{Salvadori2007} indicated that the threshold of metallicity for the formation of low-mass stars should be $\sim 10^{-4}Z_{\odot}$ in order to reproduce the metallicity distribution of observed metal-poor stars in the Galactic halo.
This value does not coincide with the theoretical expectation for the critical metallicity for dust-induced fragmentation, $Z_{\rm crit} = 10^{-6} -10^{-5}Z_{\odot}$.

This discrepancy can be alleviated by reducing the dust depletion factor in low-metallicity environments.
Namely, if the dust depletion factor is smaller than in the solar neighborhood, 
$\mathcal D$ may still fall short of $\mathcal D_{\rm crit}$
even for the gas with nominal metallicity higher than $Z_{\rm crit}$.

One way to reduce $f_{\rm dep}$ is slower growth of dust mass in low-metallicity interstellar medium \citep[e.g.,][]{Asano2013}.
Recently \citet{Remy-Ruyer2014} suggested that the observed lower dust depletion factor in the low-metallicity galaxies in the local universe could be explained by such mechanism.

Another possible mechanism, which we focus on in this paper, is the dust evacuation from star-forming regions caused by radiation feedback from massive stars.
In this case, dust grains are pushed out due to the radiation force and can be decoupled from the gas component.
In fact, the lower amount of dust grains in the observed H{\sc ii} regions around massive stars than in the H{\sc i} gas is likely to be caused by such mechanism \citep{Draine2011,Akimkin2015, Akimkin2017,Ishiki2018}.
Also, the radiation force has been claimed to be able to expel dust grains even from galactic haloes \citep{Chiao1972,Ferra1991}.
Similarly, we speculate that the dust evacuation also occurs in low-metallicity star-forming clouds in the early universe, thereby allowing massive star formation to continue even with certain metal enrichment.

In this paper, we study the condition for the formation of dust-free star-forming clouds as a result of dust evacuation by the radiation force. 
We have found that the dust evacuation successfully occurs in clouds with high column density $N_{\rm H} \simeq 10^{24} - 10^{26}~{\rm cm^{-2}}$ and very low metallicity $Z \la 10^{-2}~Z_{\odot}$.
In terms of galaxy properties, this occurs more easily for higher halo mass and formation redshift, e.g., halos of $\sim 10^{9}~M_{\odot}$ ( $\sim 10^{7}~M_{\odot}$ ) at $z \sim 3$ ( $z\sim 9$, respectively ), as long as very low-metallicity gas is available.

We organize the rest of the paper as follows.
In Section \ref{dust_evac1}, we first consider the dust evacuation from a homogeneous spherical star-forming cloud.
We then estimate the conditions for the dust evacuation from a galactic disk in Section  \ref{dustevac2}.
In Section \ref{suppression}, we investigate the possible effects that inhibit the dust evacuation.
Finally, we summarize our results and give discussions in Section \ref{matome}.
The effect of the Coulomb drag force on the terminal velocity of grains is described in Appendix \ref{apd1}.

\section{Dust grain Evacuation from a Star Forming Cloud}\label{dust_evac1}

\subsection{Formation efficiency and radiation emissivity of star clusters}
\label{eps_and_strformationrate}
We first consider the simplest case of spherical and uniform density star-forming clouds. 
For a cloud with the mass $M_{\rm cl}$ , radius $R_{\rm cl}$ and 
star formation efficiency (SFE) $\epsilon_{*}$, 
the total stellar mass $M_{*}$ formed in the cloud is given by 
\begin{eqnarray}
	M_{*} = \epsilon_{*} M_{\rm cl} = 10^{5}~ M_{\odot} \left( \frac{\epsilon_{*}}{0.1} \right)  \left( \frac{M_{\rm cl}}{10^6 ~M_{\odot}} \right). \label{0723.1}
\end{eqnarray}

\begin{figure}
	\begin{center}
		\includegraphics[width=\columnwidth]{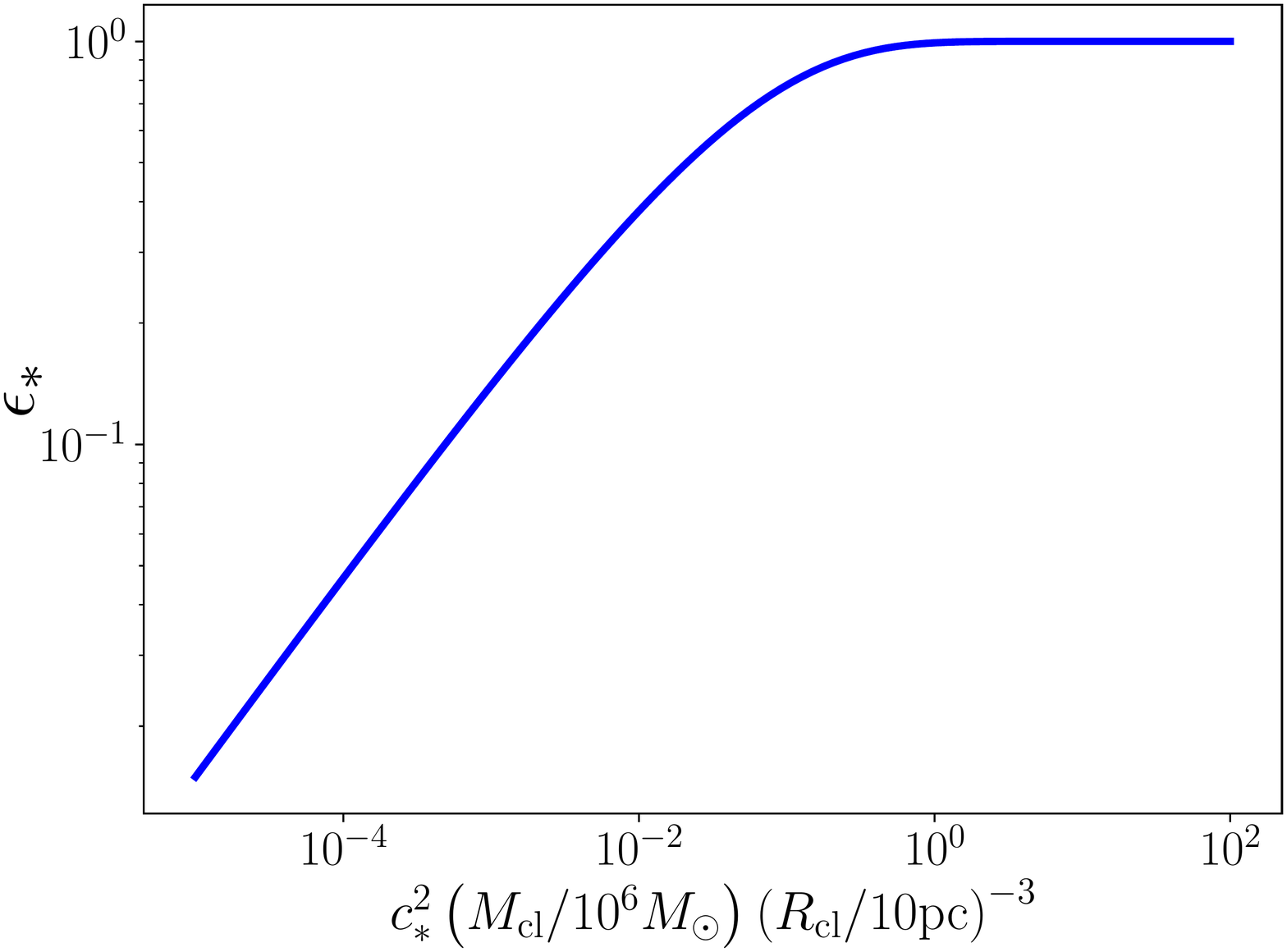}
	\end{center}
\caption{The SFE $\epsilon_{*}$ as a function of the cloud mass $M_{\rm cl}$, the cloud radius $R_{\rm cl}$ and $c_{*}$ ( Eq. \ref{0723.4}). 
We used $t_{\rm OB} = 2.5 \times 10^{6} {\rm yr}$.}
\label{fig0806.1}
\end{figure}

The SFE $\epsilon_{*}$ is likely to depend on the properties of star-forming clouds \citep{Katz1992}. Here we estimate the growth rate of the total stellar mass based on the local free-fall time $t_{\rm ff}$ as
\begin{eqnarray}
	\frac{d M_{*}}{d t} = c_{*} \frac{M_{\rm cl} - M_{*}}{t_{\rm ff}}, \label{0723.2}
\end{eqnarray}
where $c_{*}$ is a parameter characterizing the star formation rate (SFR), and $t_{\rm ff}$ is defined as 
\begin{eqnarray}
	t_{\rm ff} = \sqrt{\frac{3 \pi}{32 G m_{\rm H} n_{\rm H}}}, \label{0907.1}
\end{eqnarray}
where the number density of gas $n_{\rm H}$ is given as
\begin{eqnarray}
	n_{\rm H} = \frac{M_{\rm cl}}{\frac{4}{3} \pi R_{\rm cl}^{3} m_{\rm H}} = 9.7 \times 10^{3} {\rm cm^{-3}}  \left( \frac{R_{\rm cl}}{10~ {\rm pc}} \right)^{-3} \left( \frac{M_{\rm cl}}{10^{6}~M_{\odot}} \right). \label{0729.1}
\end{eqnarray}
We estimate the total stellar mass $M_{*}$ by integrating Equation \eqref{0723.2} with the condition $M_{*} = 0$ at $t=0$:
\begin{eqnarray}
	M_{*} = M_{\rm cl} \left[ 1 - \exp \left(- c_{*} \frac{t}{t_{\rm ff}} \right) \right]. \label{0723.3}
\end{eqnarray}
Massive stars are expected to end their lives as SNe.
Their feedback likely destroys the star-forming clouds and stops further star formation. 
In addition, newly formed dust grains will be supplied in those events.
Therefore, we use the SFE at the lifetime of OB stars, $t = t_{\rm OB}$, in estimating the condition for the dust evacuation:
\begin{eqnarray}
	\epsilon_{*} = \left[ 1 - \exp \left(- c_{*} \frac{t_{\rm OB}}{t_{\rm ff}} \right) \right]. \label{0723.4}
\end{eqnarray}
The relation among $\epsilon_{*}$, $M_{\rm cl}$, $R_{\rm cl}$ and $c_{*}$ is shown in Figure \ref{fig0806.1}.
For $M_{\rm cl} = 10^{6}~M_{\odot}$, $R_{\rm cl} = 10~{\rm pc}$, 
the SFE becomes $\epsilon_{*} = 0.38$ ($4.7 \times 10^{-2}$) for $c_{*} = 0.1$ ($0.01$, respectively).

We use the IMF $\, \psi (m_{*}) \,$ of \citet{Larson1998}: 
\begin{eqnarray}
	\psi (m_{*})= \frac{dN}{d \log m_{*}} \propto \left(1 + \frac{ m_{*} }{ m_{\rm ch}} \right)^{-1.35}, \label{3.1.1}
\end{eqnarray}
where $m_{\rm ch}$ is the characteristic stellar mass formed.
 In this paper, we consider the case of $m_{\rm ch} = 10~M_{\odot}$ as the fiducial one, 
 which is claimed by \citet{Komiya2007} based on the carbon enhanced metal-poor (CEMP) star statistics in the Galaxy.
We also study the cases with low stellar mass $m_{\rm ch} = 1~M_{\odot}$ as in the solar neighborhood \citep[e.g.,][]{Kroupa2001}, and 
very massive stars $m_{\rm ch} = 50~M_{\odot}$ as in the primordial star formation \citep{Hosokawa2011}.  
We take the mass range from $0.1~M_{\odot}$ to $300~M_{\odot}$ in all the cases.

\begin{figure}
	\begin{center}
		\includegraphics[width=\columnwidth]{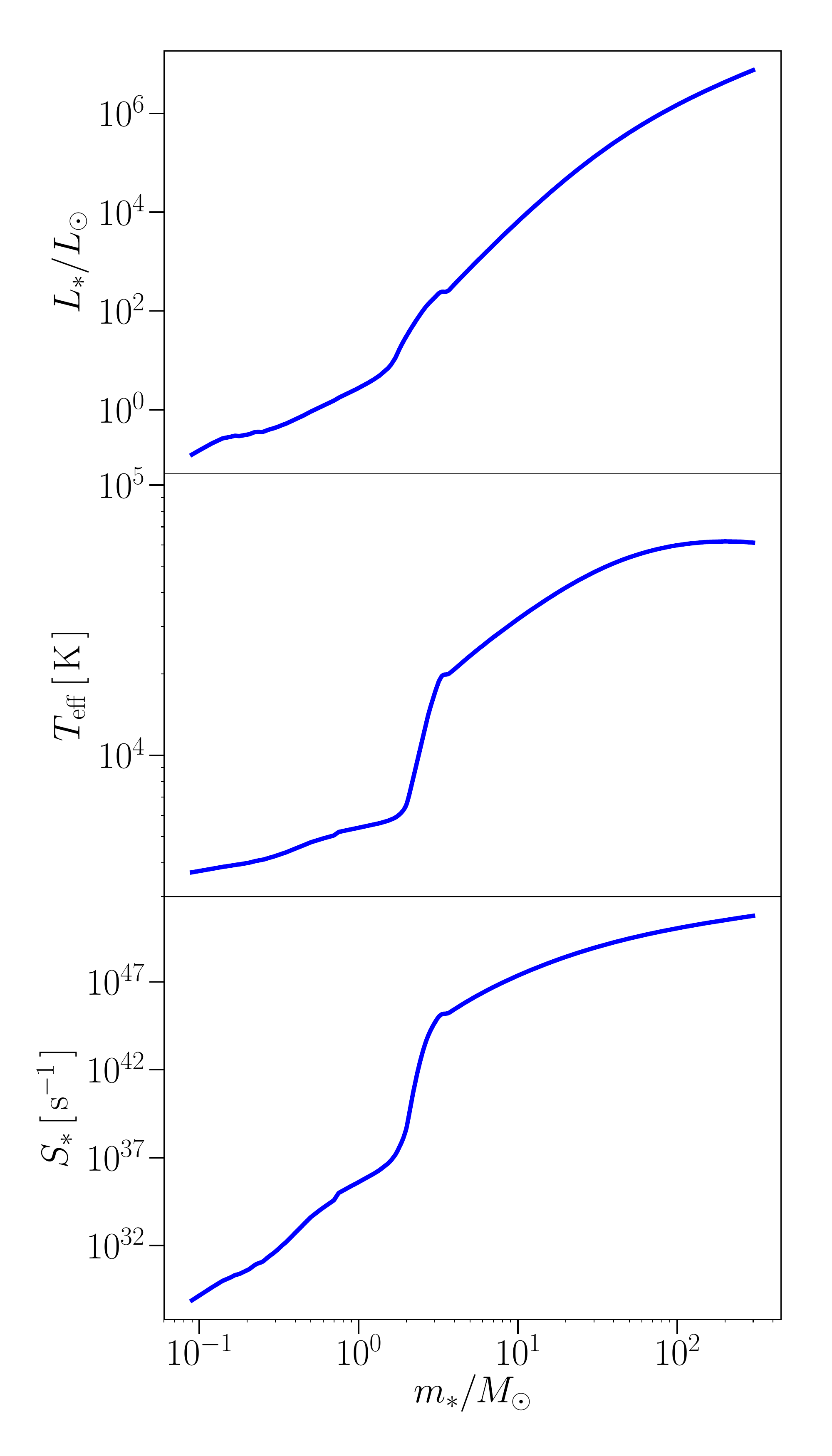}
	\end{center}
\caption{The luminosity $L_{*}$ (top), the effective temperature $T_{\rm eff}$ (middle) and ionizing photon emissivity $S_{*}$ (bottom) of stars at metallicity $10^{-2}~Z_{\odot}$ and  $t = 10^{6}~{\rm yr}$ as a function of their mass \citep{Chen2015}. \label{zu3.1}}
\end{figure}

Next, we estimate the total luminosity $L_{\rm tot}$ and the total ionizing photon emissivity $S_{\rm tot}$ of a star cluster.
We use the stellar isochrone calculated by \citet{Chen2015} with metallicity $Z =10^{-2}~Z_{\odot}$ and at $t = 10^{6}~{\rm yr}$ 
roughly corresponding to half the average cloud lifetime.
Figure \ref{zu3.1} illustrates the luminosity $L_{*}(m_{*})$, the effective temperature $T_{\rm eff}(m_{*})$, 
and the ionizing photon emissivity $S_{*}$ as functions of the stellar mass.
The luminosity and the ionizing photon emissivity per 
unit stellar mass are calculated by taking the average with the weight of the IMF (Eq. \ref{3.1.1}).
They are presented in Table \ref{tabal0729.1}.
 \begin{table}
  \caption{The luminosity and the ionizing photon emissivity per unit stellar mass for different characteristic stellar mass $m_{\rm ch}$. 
  The numbers are in unit of $( L_{\odot} ~ M_{\odot}^{-1} )$ and $( s^{-1}~M_{\odot}^{-1} )$.}
  \label{tabal0729.1}
  \centering
  \begin{tabular}{ccc}
    \hline \hline
     $m_{\rm ch} \, [\, M_{\odot} \,]$  & $\langle L_{*} /  m_{*}  \rangle $ & $\langle S_{*} / m_{*} \rangle$  \\ 
    \hline
    $1 $  & $2.9 \times 10^{3}$ & $2.2 \times 10^{47} $ \\
    $10 $  & $6.7 \times 10^{3}$ & $5.0 \times 10^{47} $ \\
    $50 $  & $1.1 \times 10^{4}$ & $8.3 \times 10^{47} $ \\
    \hline
  \end{tabular}
 \end{table}
In the fiducial case with $m_{\rm ch} = 10~M_{\odot}$, the total luminosity and ionization emissivity are
\begin{eqnarray}
	&& L _{\rm tot} = 6.7 \times 10^{8} \, L _{\odot} \, \left( \frac{\epsilon_{*}}{0.1} \right)  \left( \frac{M_{\rm cl}} {10^{6}M_{\odot}} \right),  \label{3.1.5} 
\end{eqnarray}
and
\begin{eqnarray}
&& S_{\rm tot}  = 5.0 \times 10^{52} \, {\rm s^{-1}} \,   \left( \frac{\epsilon_{*}}{0.1} \right) \left( \frac{M_{\rm cl}} {10^{6}M_{\odot}} \right). \label{3.1.6}
\end{eqnarray}

\subsection{Dust evacuation time}\label{sec.det}
\begin{figure}
    \begin{center}
      \includegraphics[width=\columnwidth]{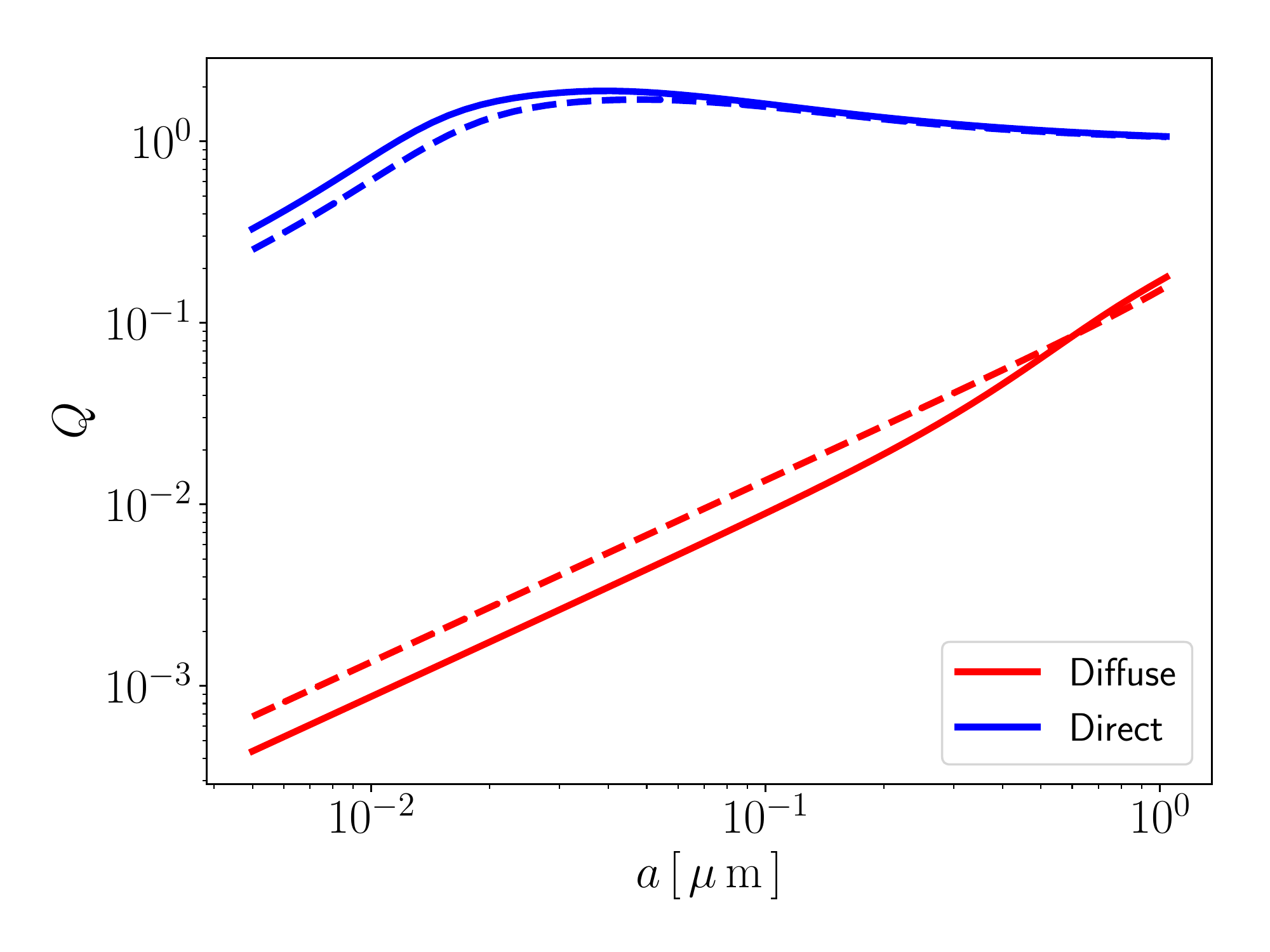}
    \end{center}
    \caption{ 
     The absorption efficiency factor $Q$ to geometric area as a function of dust size. 
     Blue and red lines show the values to the stellar radiation of black body with the effective temperature of $2 \times 10^{4}~\rm K$ and the diffuse thermal emission of dust with the temperature of $100~\rm K$, respectively. 
    The solid- and dashed-lines represent graphite and silicate dust grains.
    We calculate the efficiency factor $Q$ from Mie theory with the dielectric function of \citet{Draine1984} and \citet{Draine2003}.  \label{zu0807.6}
    }
\end{figure}
The motion of dust grains is governed by three forces:
radiation force $F_{\rm rad}$, gravity $F_{\rm grav}$ and drag force $F_{\rm drag}$.
We estimate the dust terminal velocity $v_{\rm d}$ as follows.

Radiation force $F_{\rm rad}$ exerted on a dust grain is
\begin{eqnarray}
	F_{\rm rad} = \frac{\pi a^{2} Q L_{\rm tot}}{4 \pi r^{2} c}, \label{0915.1}
\end{eqnarray}
where $a$ is the radius of the dust grain and $Q$ is the absorption efficiency factor relative to the geometrical cross section.
The efficiency factor $Q$ depends on the size of dust grains and the frequency of radiation.
When the cloud is optically thin, radiation field inside the cloud is dominated by the direct light 
from massive stars, which can be approximated by the black-body spectrum of $\sim 2 \times 10^{4} ~{\rm K}$.
On the other hand, in the optically thick case the stellar light is absorbed and re-emitted by dust grains, 
and the radiation is dominated by the the diffuse light, which is roughly the black body of $\sim 100 ~{\rm K}$ \citep[e.g.,][]{Fukushima2017}. 
Figure \ref{zu0807.6} shows
the size dependence of the efficiency factor $Q$ of graphite and silicate grains both for the direct and diffuse light.
Note $Q \simeq 1$ ($\simeq 10^{-2}$) for the direct (diffuse, respectively) light for grains with $a \simeq 0.1~\mu {\rm m}$.

The ratio of radiation force to the gravity on a dust grain is
\begin{align}
  \frac{F_{\rm rad}}{F_{\rm grav}} &= \frac{\pi a^2 Q L_{\rm tot}}{4 \pi c G M_{\rm cl} m_{\rm d}} \nonumber \\
   &= 1.3 \times 10^{3} \left( \frac{Q}{1} \right)\left( \frac{\rho_{\rm d}}{3 {\rm \hspace{1mm} g \hspace{1mm} cm^{-3}}  } \right)^{-1} \left( \frac{a}{0.1\mu {\rm m}} \right)^{-1}  \left( \frac{\epsilon_{*}}{0.1} \right).  \label{2.1.4}
\end{align} 
That is, the gravity is negligible compared to the radiation force.
The terminal velocity of dust $v_{\rm d}$ is determined by the balance between radiation and drag forces alone.

We consider the collisional gas drag $F_{\rm collision}$ and the Coulomb drag $F_{\rm Coulomb}$ as the  
drag force on dust grains, $F_{\rm drag}=F_{\rm collision}+F_{\rm Coulomb}$ (see Appendix \ref{apd1} for the expression).
The Coulomb drag and thus the relative velocity of dust grains sensitively depend on the ionization degree $x_{\rm e}$ in the cloud (for the estimation of the terminal velocity, see Appendix  \ref{apd1}). 
Figure \ref{zu0907.1} shows the terminal velocity of a dust grain with $0.1~\mu {\rm m}$ 
as a function of ionization degree for the cloud with $M_{\rm cl} = 10^{6}~M_{\odot}$, $R_{\rm cl} = 10~{\rm pc}$ and $\epsilon_{*} = 0.1$.
\begin{figure}
    \begin{center}
      \includegraphics[width=\columnwidth]{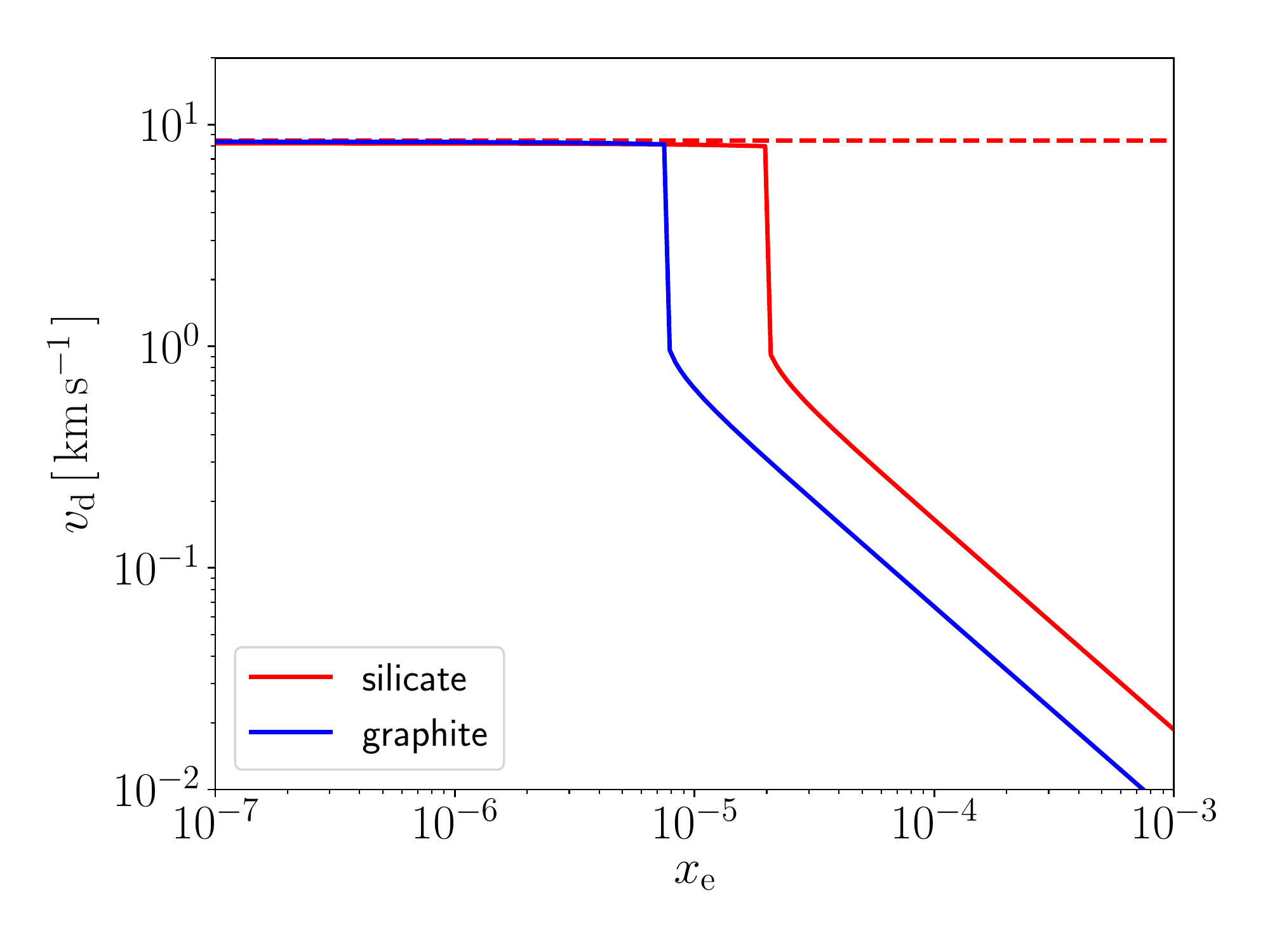}
    \end{center}
    \caption{ 
    The terminal velocity of dust grains in a star-forming cloud $M_{\rm cl} = 10^{6}~ M_{\odot}$, $R_{\rm cl} = 10 ~ {\rm pc}$ and $\epsilon_{*} = 0.1$. 
    The solid and dashed lines show the velocities estimated with and without Coulomb drag force, respectively. 
    The blue and red lines are for the graphite and silicate grains, respectively. 
 \label{zu0907.1}}
\end{figure}
The solid lines show the terminal velocities calculated by considering both the collisional and Coulomb drag forces, while 
the dashed lines are those with the collisional drag force alone.
For $x_{\rm e} \ga 10^{-5}$, the Coulomb drag dominates the collisional drag, and the velocity of dust grains suddenly falls below $1 ~{\rm km \, s^{-1}}$.
In contrast, the Coulomb drag does not work for $x_{\rm e} \la 10^{-5}$, and the terminal velocity is $\sim 10 ~ {\rm km \, s^{-1}}$, set by the collisional drag.

In our model, free electrons are supplied by the ionization either of heavy elements by far ultraviolet (FUV) light 
or of hydrogen by cosmic rays.
The typical number density $n(\rm M^{+})$ of heavy elements ionized by FUV photons of $h \nu < 13.6 \rm{e V}$ in H{\sc i} regions is estimated by \citet{Draine2011a} as
\begin{eqnarray}
x_{\rm M} = n({\rm M^{+}}) / n_{\rm H} \simeq 1.1 \times 10^{-4}  \label{0709.13} 
\end{eqnarray}
for the solar metallicity gas.
Therefore, in the low-metallicity case of $Z \la 10^{-1}Z_{\odot}$, we expect that $x_{\rm M} \la 10^{-5}$ and 
thus the Coulomb drag can be neglected.
The hydrogen ionization fraction by cosmic rays is calculated by the balance between the ionization rate of cosmic rays $\xi_{\rm CR}$ and the recombination rate of ionized hydrogen,
\begin{eqnarray}
	n_{\rm H} \xi_{\rm CR} 
= \alpha_{\rm B} n_{\rm H}^{2} x_{\rm e}^{2}, \label{0802.1}
\end{eqnarray}
where $\alpha_{\rm B}$ is the case-B recombination coefficient.
Thus, $x_{\rm e}$ is given as, 
\begin{eqnarray}
	x_{\rm e} &=& \left( \frac{\xi_{\rm CR}}{\alpha_{\rm B} n_{\rm H}} \right)^{1/2} \nonumber \\
	& =  & 3.8 \times 10^{-5} \left(\frac{\xi_{\rm CR}}{10^{-16} {\rm s^{-1}} } \right)^{1/2}  \left( \frac{R_{\rm cl}}{10 ~ \rm{pc}} \right)^{3/2} \left( \frac{M_{\rm cl}}{10^{6} ~M_{\odot}} \right)^{-1/2} , \nonumber  \\ \label{0802.2}
\end{eqnarray}
where we have used $\alpha_{\rm B} = 7 \times 10^{-12} ~{\rm cm^{3} \, s^{-1} }$ at $ T = 100 ~ {\rm K}$ \citep{Hummer1987}.
Ionized hydrogen also recombines via the charge exchange reactions with heavy elements (e.g. CO) 
and the ionization fraction becomes smaller than $10^{-6}$ for number density higher than $\sim 10^{3} ~{\rm cm^{-3}}$\citep{McKee1989}.
Thus Equation \eqref{0802.2} gives the upper limit of ionization degree, which is likely to be smaller 
than the critical value above which the Coulomb drag becomes important.
Therefore, we ignore the Coulomb drag below and estimate the dust terminal velocity from the balance between the collisional drag and radiation forces.
Here, we only consider the cosmic rays as the ionization source. 
In section \ref{sec4.1}, we discuss the effects of other sources, such as X-rays or mixing of ionized and neutral gases.

For supersonic relative velocity between the gas and dust, the collisional drag force can be written as 
$F_{\rm collision}= \pi a^2v_{\rm d}^2 n_{\rm H} m_{\rm H}$  \citep{DS1979}.
The terminal velocity $v_{\rm d}$ is then given as
\begin{eqnarray}
  v_{\rm d} &=& \sqrt{\frac{L_{\rm tot} Q}{4 \pi R_{\rm cl}^2 c n_{\rm H} m_{\rm H}}} \nonumber \\
  &=& 6.7 ~  {\rm km \hspace{1mm}  s^{-1}}  \left( \frac{Q}{1} \right)^{1/2} \left( \frac{\epsilon_{*}}{0.1} \right)^{1/2} 
  \left( \frac{R_{\rm cl}}{10 ~ {\rm pc}} \right)^{1/2}. \label{1.6.3}
\end{eqnarray}
Dust evacuation time $t_{\rm evac}$ from a star-forming cloud is estimated as
\begin{eqnarray}
  t_{\rm evac} &=& \frac{R_{\rm cl}}{v_{\rm d}} \nonumber \\
  &=& \sqrt{\frac{4 \pi c n_{\rm H} m_{\rm H}}{L_{\rm tot}Q} } R_{\rm cl}^2 \nonumber \\
&=& 1.5 \times 10^{6} ~ {\rm yr}  \left( \frac{Q}{1} \right)^{-1/2} \left( \frac{\epsilon_{*}}{0.1} \right)^{-1/2}  \left( \frac{R_{\rm cl}}{10~ {\rm pc}} \right)^{1/2}.  \label{1.6.4}
\end{eqnarray}
Note that $t_{\rm evac}$ increases with the cloud radius.

Note that the drag time  $t_{\rm drag}$ required for transporting momentum from a grain to the gas is given as 
\begin{align}
	 t_{\rm drag} & = \frac{m_{\rm d} v_{\rm d}}{F_{\rm drag}} \nonumber \\
  	& = 1.2 \times 10^{2} ~{\rm yr} \left( \frac{Q}{1} \right)^{-1/2} \left( \frac{\rho_{\rm d}}{3 \,{\rm g \, cm^{-3}}} \right) \nonumber \\
	&\hspace{1cm} \times  \left( \frac{a}{0.1 \mu {\rm m}} \right) \left( \frac{\epsilon_{*}}{0.1} \right)^{-1/2} \left( \frac{R_{\rm cl}}{10 ~{\rm pc}} \right)^{5/2} \left( \frac{M_{\rm cl}}{10^{6}M_{\odot}} \right)^{-1}, \label{0807.1}
\end{align}
which is much shorter than the evacuation time $t_{\rm evac}$ (Eq. \ref{1.6.4}).
Therefore, our use of the terminal velocity as the dust velocity is justified.

In estimating $t_{\rm evac}$, we have used the efficiency factor of $Q=1$ corresponding 
to the optically thin case. 
If the cloud is optically thick, on the other hand, the radiation force on to dust grains is imparted mostly 
by the diffuse light and the efficiency factor drops to $Q=10^{-2}$.
With such low efficiency factor, the dust evacuation time (Eq. \ref{1.6.4}) becomes as long as 
$t_{\rm evac} = 1.5 \times 10^{7}{\rm yr}$, exceeding the lifetime of massive stars $t_{\rm OB}$.
Thus, optically-thick clouds would be destructed or replenished with newly formed dust grains 
due to the SN feedback before dust grains are evacuated.
Therefore, the clouds must be optically thin initially to become dust-free by the dust evacuation .
Using the dust opacity $\kappa = 350~ {\rm cm^{2} \hspace{1mm} g^{-1}} \left( Z / Z_{\odot} \right)$ 
at the wavelength $\lambda _{\rm max} = 0.254 \rm \mu m$,
which is the peak of the black body spectrum of $T = 2 \times 10^{4}~{\rm K}$,
we estimate the optical depth of the cloud as
\begin{eqnarray}
	\tau &=& \rho \kappa R_{\rm cl} \nonumber \\
	&=& 175 \left( \frac{\kappa}{350 \, {\rm cm^{2} g^{-1}}} \right)  \left( \frac{R_{\rm cl}}{10 ~ \rm{pc}} \right)^{-2} \left( \frac{M_{\rm cl}}{10^{6}M_{\odot}} \right) \left( \frac{Z}{Z_{\odot}} \right). \label{2.1.5}
\end{eqnarray}
The condition for optical thinness thus corresponds to metallicity
\begin{eqnarray}
Z < 5.7 \times 10^{-3} Z_{\odot}  \left( \frac{\kappa}{350 \, {\rm cm^{2} g^{-1}}} \right)^{-1} \left( \frac{R_{\rm cl}}{10 ~ \rm{pc}} \right)^{2} \left( \frac{M_{\rm cl}}{10^{6}M_{\odot}} \right)^{-1} .\label{0726.1}
\end{eqnarray}

In the estimation above, we do not consider the dust size dependence 
of the efficiency factor and always use the value $Q=1$. 
As seen in Figure \ref{zu0807.6}, this is only true for grains larger than $\sim 0.01\mu {\rm m}$.
For grains smaller than $0.01\mu {\rm m}$, the efficiency factor becomes $Q \lesssim 0.3$, 
which makes the evacuation time longer due to the weaker radiation force.
In fact, by substituting $Q = 0.3$ into Equation \eqref{1.6.4}, dust evacuation time becomes $t_{\rm evac} = 2.7 \times 10^{6}~{\rm yr}$, slightly exceeding $t_{\rm OB}$.
In this case, small grains with size $a<0.01 ~ \mu {\rm m}$ would remain in the cloud.
In Section \ref{sec4.2}, we additionally discuss cases in which small dust grains are dominant.
The dependence of timescales on the grain size is discussed in Appendix  \ref{apd1}.

\subsection{Cloud lifetime}\label{cloud_dest}
For star formation from the dust-free gas, dust evacuation time must to be shorter than destruction time of the cloud.
Lifetime of the star-forming cloud $t_{\rm cl}$ is estimated with
the timescale that either type-II SNe occur $t_{\rm OB}$ or the ionizing front reaches the boundary of the cloud $t_{\rm HII}$:
\begin{eqnarray}
	t_{\rm cl} = {\rm min} (t_{\rm OB}, t_{\rm HII}). \label{0907.2}
\end{eqnarray}
The timescale of SNe is determined by the lifetime of massive stars, 
and we use $t_{\rm OB} = 2.5\times 10^{6}~\rm{yr}$, which is the lifetime for $m_{*} = 120 M_{\odot}$ \citep{Schaerer2002}, in this work.

The expansion time of an H{\sc ii} region $t_{\rm HII}$ can be calculated as follows.
The ionizing front reaches the Str\"omgren radius
\begin{eqnarray}
  R_{\rm S0} &=& \left( \frac{3 {S_{\rm tot}}}{4 \pi n_{\rm H}^2 \alpha_{\rm B}} \right)^{1/3} \nonumber \\
   &=& 2.5  {\rm pc} \left( \frac{\epsilon_{*}}{0.1} \right)^{1/3} \left( \frac{R_{\rm cl}}{10 ~ \rm{pc}} \right)^{2}   \left( \frac{M_{\rm cl}}{10^6 M_{\odot}} \right)^{-1/3},   \label{1.6.10}
\end{eqnarray}
almost instantly in a few hundred years or so.
Here, we have used Equation \eqref{3.1.6} for number density of gas and case-B recombination rate $\alpha_{\rm B} = 2.6 \times 10^{-13} {\rm cm^{3} \, s^{-1}}$ \citep[at $T=10^{4}~\rm K$, ][]{Osterbrock2006}, 
Thereafter the pressure imbalance drives the further expansion of the H{\sc ii} region.
The radius of the H{\sc ii} region in this phase is given by \citep{Spitzer1978}
\begin{eqnarray}
  R_{\rm HII} = R_{\rm S0} \left[1 + \frac{7}{4} \frac{c_{\rm HII} t}{R_{\rm S0}} \right]^{4/7}. \label{1.6.11}
\end{eqnarray}
By equating $R_{\rm HII}$ with the cloud radius $R_{\rm cl}$, we obtain  
the time for H{\sc ii} region to reach the cloud boundary:
\begin{eqnarray}
  t_{\rm HII} &=& \frac{4}{7} \frac{R_{\rm S0}}{c_{\rm HII}} \left[ \left( \frac{R_{\rm cl}}{R_{\rm S0}} \right)^{7/4} - 1 \right] \nonumber \\
  &\simeq& 1.4 \times 10^{6} {\rm yr}  \left(\frac{\epsilon_{*}}{0.1}\right)^{-1/4}  \left(\frac{R_{\rm cl}}{10 ~{\rm pc}}\right)^{1/4}  \left(\frac{M_{\rm cl}}{10^{6}M_{\odot}}\right)^{1/4}. \hspace{5mm}  \label{1.6.12}
\end{eqnarray}
In the numerical estimate above, we have used the temperature in the H{\sc ii} region $T =10^{4}~{\rm K}$ and 
the sound speed $c_{\rm HII} = 11.4~ {\rm km \, s^{-1}}$.
Note that the expansion time of H{\sc ii} region has weak dependence on the cloud radius compared with the dust evacuation time (Eq. \ref{1.6.4}).

We now examine whether the radiation force on the dust destroys the star-forming cloud itself.
Since the ratio of radiation force to the gravity at $R_{\rm cl}$ is 
\begin{eqnarray}
  \frac{F_{\rm rad}}{F_{\rm grav}} &=& \frac{\kappa_{\rm d} L_{*}}{4 \pi c G M_{\rm cl}}  \nonumber \\
  &=&  1.8 \times 10^{-2} \left( \frac{\kappa}{350 ~ {\rm cm^{2} g^{-1}}} \right) \left( \frac{\epsilon_{*}}{0.1} \right) \left( \frac{Z}{10^{-3}Z_{\odot}} \right), \label{2.1.3}
\end{eqnarray}
the radiation force is smaller than gravity on the gas component in optically-thin clouds. 
Such clouds can avoid being disrupted by the radiation force.

\subsection{The condition for dust-free cloud formation} \label{sec.dustless}
We here obtain the condition for formation of the dust-free star-forming clouds. 
In order for stars to form from the dust-free gas, the dust evacuation time be shorter 
than the cloud lifetime $t_{\rm cl}$, and also some gas still be available at this moment 
for subsequent star formation. 

\begin{figure}
    \begin{center}
      \includegraphics[width=\columnwidth]{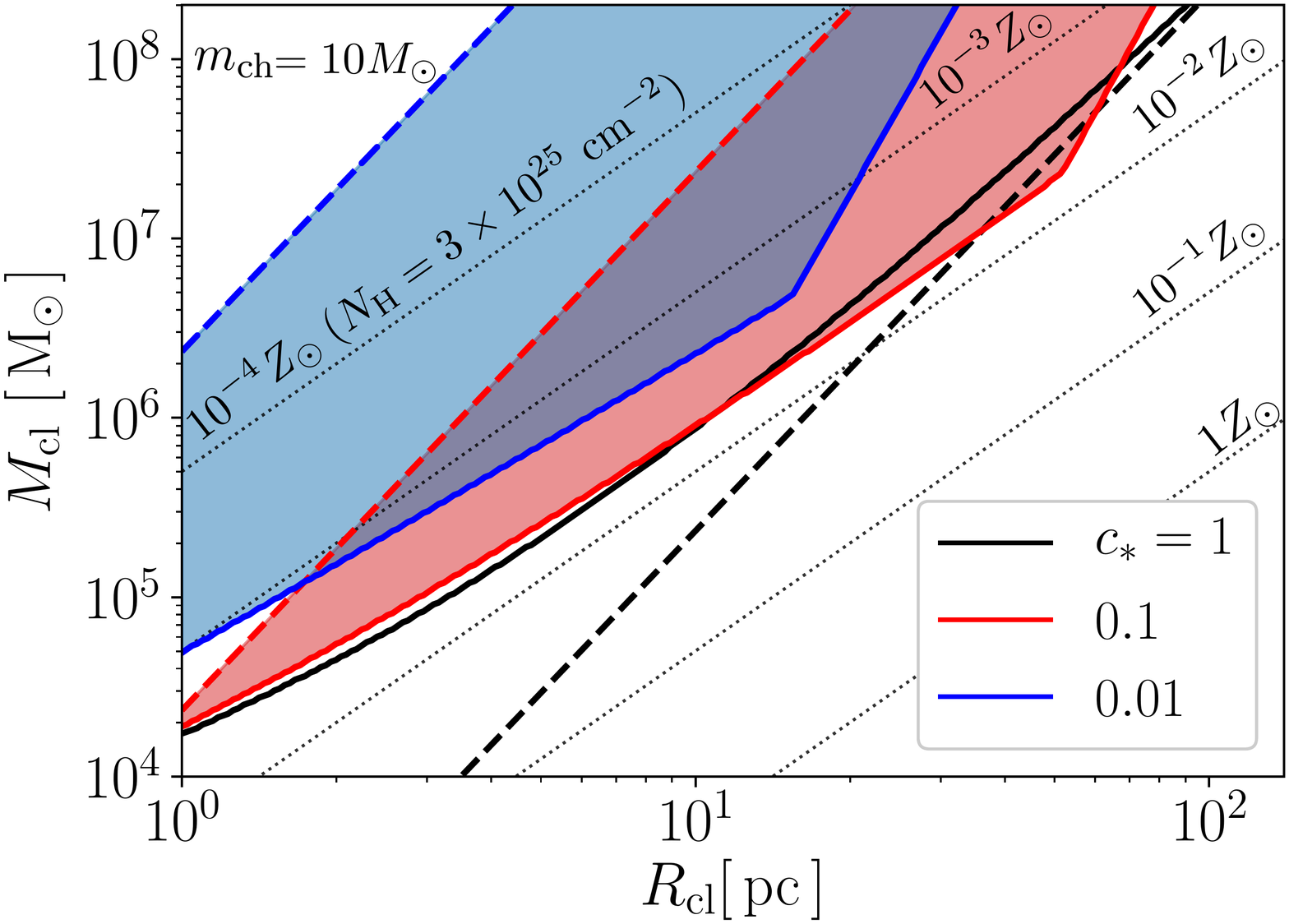}
    \end{center}
    \caption{
      The conditions for the formation of dust-free star-forming clouds in the case of $m_{\rm ch} = 10M_{\odot}$.
     The black, red, and blue solid lines denote the condition of $t_{\rm evac} < t_{\rm cl}$ 
    in the cases with $c_{*} = 1$, $0.1$ and $0.01$, respectively.
    Above the dashed line, SFE becomes $\epsilon_{*} > 0.9$.
    The blue and red shaded regions represent that the dust evacuation occurs and more than 10 per cent of initial gas remains at this epoch.
    In the case with $c_{*} = 1$, there is no such area.
    The black-dot lines show the boundaries at where the optical depth becomes $\tau=1$ at each metallicity, and clouds are optically thin below these lines. 
    The relation for $\tau = 1$ between the column density $N_{\rm H}$ and metallicity $Z$ is denoted as $N_{\rm H} = 3 \times 10^{25} ~ {\rm cm^{-2}} \left( Z / 10^{-4} Z_{\odot} \right)$.
        \label{zu0807.3}}
\end{figure}

We consider the cases with the star-formation rate parameter $c_{*} = 0.01 - 1$, 
which is related to the SFE $\epsilon_{*}$ by Equation 
(\ref{0723.4}).
Figure \ref{zu0807.3} shows
the condition for the formation of a dust-free star-forming cloud in the case with the characteristic stellar mass 
$m_{\rm ch} = 10~M_{\odot}$.
The color-shaded areas show the parameter ranges of clouds that become dust-free for different values of $c_{*}$.
The black, red and blue lines correspond to the cases with $c_{*}= 1, 0.1$, and $0.01$, respectively.
The condition of the optical depth $\tau = 1$ is also shown by the thin dotted lines for indicated metallicities, 
and the clouds are optically thin below them. 
For smaller $c_{*}$, the dust evacuation time $t_{\rm evac}$ becomes longer from Equation \eqref{1.6.4}
and the clouds must be more compact for the dust evacuation.  
The SFE $\epsilon_{\ast}$ is larger than $0.9$ 
above the dashed lines in Figure \ref{zu0807.3}.
In these regions, most of the gas has been consumed by star formation by the time of the dust evacuation.
As a result, star formation from the dust-free gas is only realized in the color-shaded areas between 
the solid and dashed-lines.
In the case with $c_{*} = 1$, 
there is no shaded area, which means that star formation from the dust-free gas does not occur in any clouds below $10^{8}~M_{\odot}$. 
In the case with $c_{*} = 0.1$, this condition roughly corresponds to the column density $ 10^{24} ~{\rm cm^{-2}} \la N_{\rm H} \la 10^{25}~{\rm cm^{-2}} $ and the metallicity $Z \la 10^{-2} ~Z_{\odot}$. 
Also, in the case with $c_{*} = 0.01$, the condition becomes  $ 3 \times 10^{24} ~{\rm cm^{-2}} \la N_{\rm H} \la 10^{26}~{\rm cm^{-2}} $ and $Z \la 10^{-3} ~Z_{\odot}$.
Namely, in compact and low-metallicity clouds, the dust evacuation is likely to occur.
Comparing this value with the column density of local giant molecular clouds (GMCs) $N_{\rm H} \simeq 10^{22}~{\rm cm^{-2}}$ \citep{Solomon1987} or OB star-forming clumps $N_{\rm H} \simeq 10^{24}~{\rm cm^{-2}}$ \citep{Plume1997}, the dust evacuation does not occur in typical GMCs, while it is satisfied in local massive star-forming clumps if the metallicity were low enough.
\begin{figure}
    \begin{center}
      \includegraphics[width=\columnwidth]{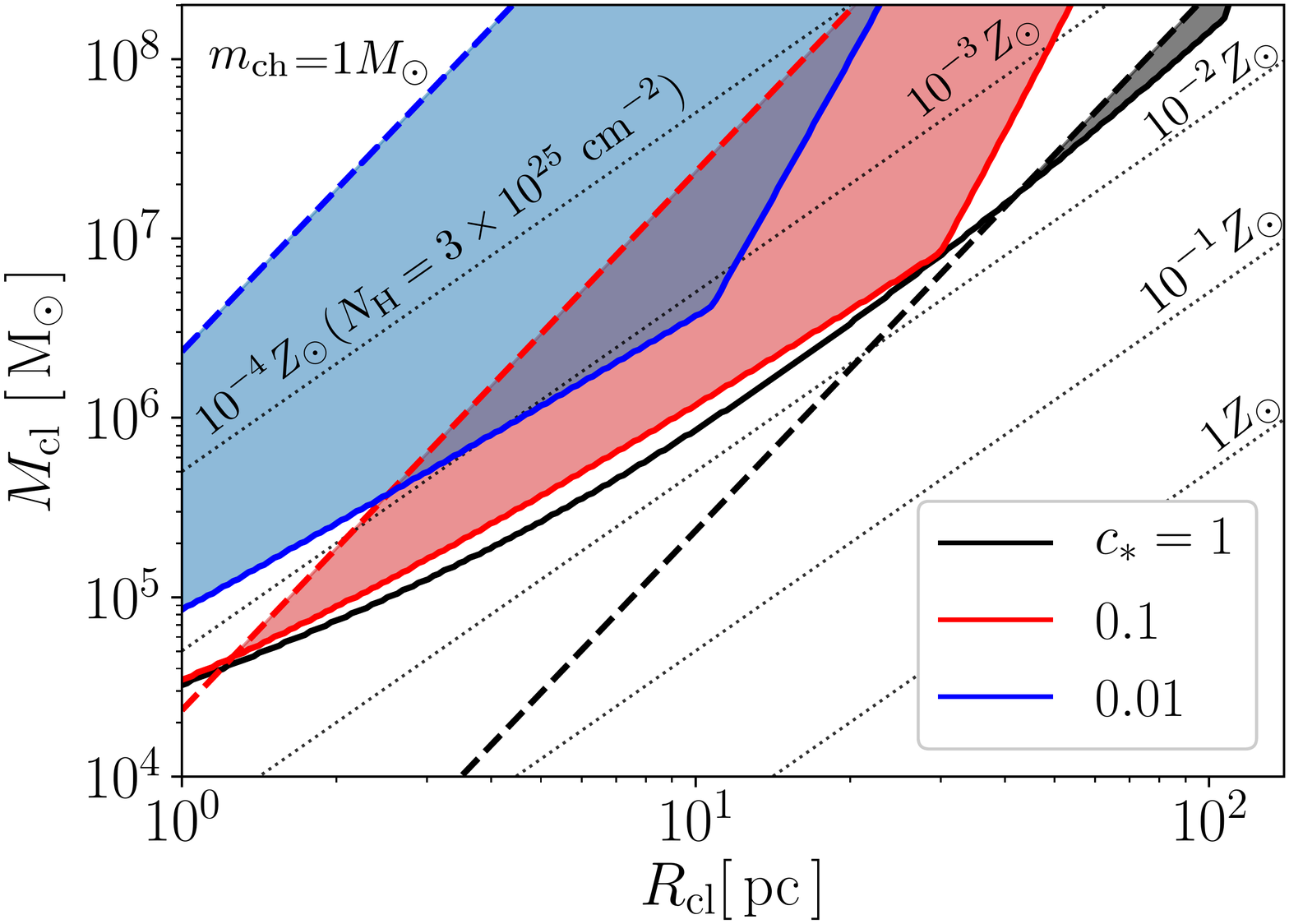}
    \end{center}
    \caption{
    Same as Figure \ref{zu0807.3}, bur for the case of lower characteristic stellar mass $m_{\rm ch} = 1 M_{\odot}$.
    In this case, the dust-free star formation condition is satisfied also for $c_{*} = 1$ in a very narrow strip with $ \ga10^{7}~M_{\odot}$.
 \label{zu0807.4}}
\end{figure}
\begin{figure}
    \begin{center}
      \includegraphics[width=\columnwidth]{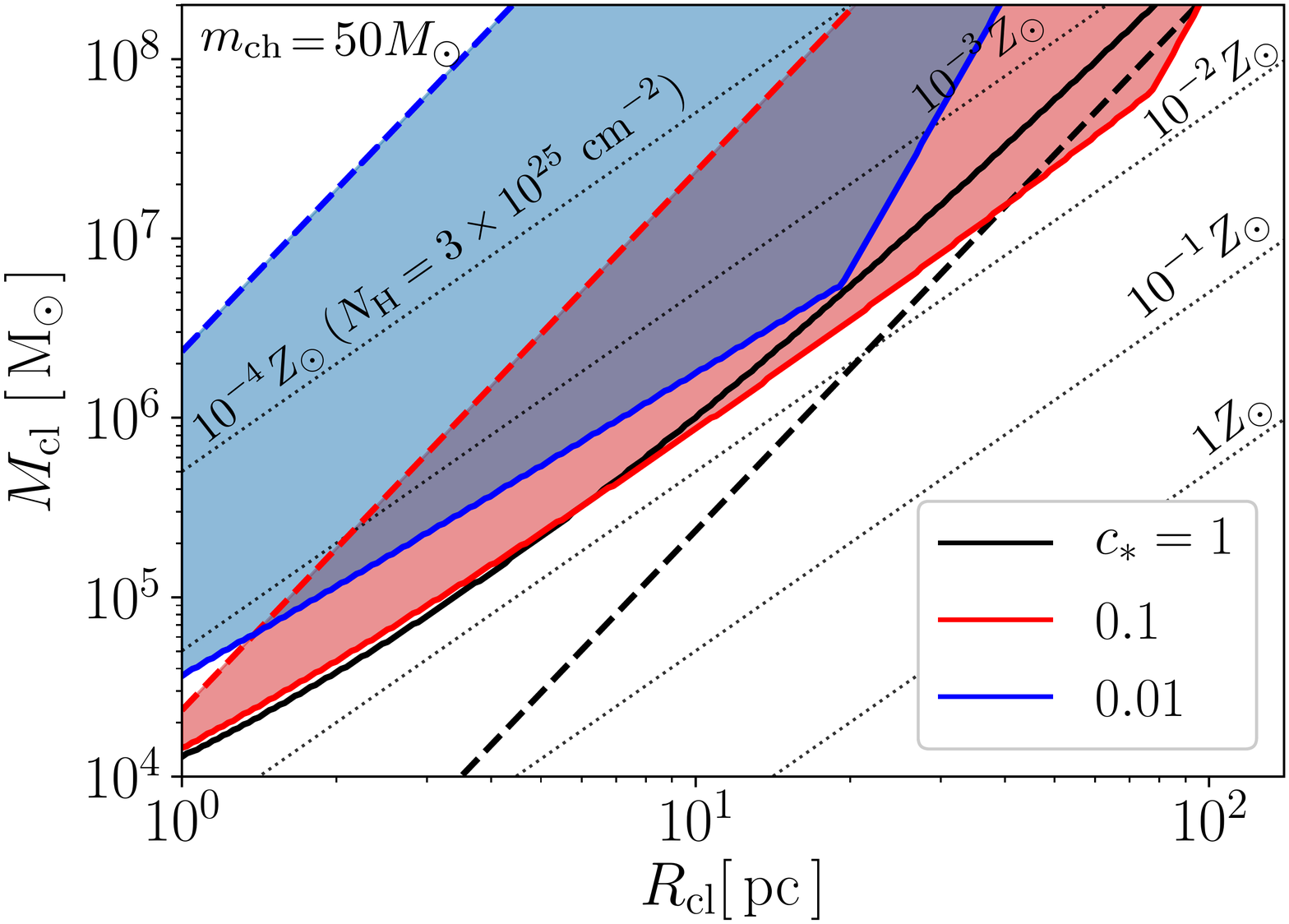}
    \end{center}
    \caption{Same as Figure \ref{zu0807.3}, but for the case of higher characteristic stellar mass 
$m_{\rm ch} = 50 M_{\odot}$. \label{zu0807.5}}
\end{figure}
The cases with different charasteristic stellar mass 
$m_{\rm ch} = 1$ and $50 M_{\odot}$ are shown in 
Figures \ref{zu0807.4} and \ref{zu0807.5}, respectively.
For higher $m_{\rm ch}$,
the luminosity per unit mass becomes higher (Table~\ref{tabal0729.1}), 
which facilitates the dust evacuation and thus results in the wider shaded regions, and vice versa.
The condition on $(M_{\rm cl}, R_{\rm cl})$ for the dust evacuation, however, does not depend so much on the value of  $m_{\rm ch}$. 

As a rule of thumb, for star formation from the dust evacuated gas to occur, the clouds should be massive or compact $N_{\rm H} \simeq 10^{24} - 10^{26}~{\rm cm^{-2}}$, and low-metallicity $Z \la 10^{-2}~Z_{\odot}$.
In addition, the SFR parameter must be somewhat smaller than unity.

\section{Dust evacuation from galactic disks}\label{dustevac2}
We have studied the dust evacuation from spherical star-forming clouds.
In this section, by applying the model developed in Section \ref{dust_evac1}, 
we study the dust evacuation from disk-like structure likely to form in low-metallicity galaxies 
in the early universe.

\subsection{Galactic disk model}
We consider a halo with mass $M_{\rm h}$ virializing at redshift $z_{\rm vir}$ and a rotationally-supported disk within it \citep{Mo1998,Oh2002}.
The virial radius is given by 
\begin{eqnarray}
	R_{\rm vir} = 2.2 {\rm kpc} \left( \frac{1 + z_{\rm vir}}{10} \right)^{-1} \left( \frac{M_{\rm h}}{10^{9}M_{\odot}} \right)^{1/3}. \label{0814.3}
\end{eqnarray}
Using the spin parameter  $\lambda = J | E | ^{1/2} / G M_{\rm h}^{5/2}$, where $E$ and $J$ are the total energy and the total angular momentum of the halo, 
the disk radius is estimated as
\begin{eqnarray}
	R_{\rm d} = \frac{\lambda}{\sqrt{2}} R_{\rm vir} = 77 {\rm pc} \left( \frac{\lambda}{0.05} \right) \left( \frac{1+z_{\rm vir}}{10} \right)^{-1} \left( \frac{M_{\rm h}}{10^{9}M_{\odot}} \right)^{1/3}. \label{0816.1}
\end{eqnarray}
The number density of the gas at radius $r$ and vertical height $z$ is 
\begin{eqnarray}
	n(r,z) = n_{0} \exp \left( - \frac{2 r}{R_{\rm d}} \right) {\rm sech} ^{2} \left( \frac{z}{\sqrt{2} z_{0}} \right), \label{0814.1}
\end{eqnarray}
where $z_{0}$ is the vertical scale height,
\begin{eqnarray}
z_{0} = \frac{c_{\rm s}}{4 \pi G \mu m_{\rm H} n_{0}} e^{r/R_{\rm d}}. \label{0814.2}
\end{eqnarray} 
We set the mean molecular weight $\mu = 1$.
Note that the disk mass can be written as 
\begin{eqnarray}
M_{\rm d} = \int dz \int 2 \pi r dr \mu m_{\rm H} n(r, z) = f_{\rm d} (\Omega_{\rm b}/ \Omega_{\rm m}) M_{\rm h},
\end{eqnarray}
where $f_{\rm d}$ is the baryon mass function taken into the disk.
Substituting Equations \eqref{0814.3}, \eqref{0816.1}, \eqref{0814.1}, and \eqref{0814.2} into the above, 
we obtain number density at the center of the disk
\begin{align}
	n_{0} = 2.0 \times 10^{4} {\rm cm^{-3}}  \left( \frac{f_{\rm d}}{0.5} \right)^{2} & \left( \frac{\lambda}{0.05} \right)^{-4}  \left( \frac{1+z_{\rm vir}}{10} \right)^{4} \nonumber \\
	& \times \left( \frac{T}{8000~ {\rm K}} \right)^{-1} \left( \frac{M_{\rm h}}{10^{9}~M_{\odot}} \right)^{2/3}.  \label{0816.2}
\end{align}
From Equation \eqref{0814.2}, the scale height of the disk is 
\begin{align}
	z_{0} = 1.6 \, {\rm pc} \, \left( \frac{f_{\rm d}}{0.5} \right)^{-1} & \left( \frac{\lambda}{0.05} \right)^{2} \left( \frac{1+z_{\rm vir}}{10} \right)^{-2}  \nonumber \\
	& \times  \left( \frac{T}{8000 {\rm K}} \right)  \left( \frac{M_{\rm h}}{10^{9}M_{\odot}} \right)^{-1/3} \exp \left( \frac{r}{R_{\rm d}} \right), \label{0816.3}
\end{align}
and the disk surface density is
\begin{align}
	\Sigma_{\rm gas} (r) &= \int dz \mu m_{\rm H} n (r, z) \nonumber \\
		    &= 0.46 \, {\rm g \, cm^{-2}} \, \left( \frac{f_{\rm d}}{0.5} \right) \left( \frac{\lambda}{0.05} \right)^{-2} \left( \frac{1+z_{\rm vir}}{10} \right)^{2} \nonumber \\
		     & \hspace{2.5cm}  \times \left( \frac{M_{\rm h}}{10^{9}M_{\odot}} \right)^{1/3} \exp \left( - \frac{r}{R_{\rm d}} \right) . \label{0816.4}
\end{align}

\begin{figure}
    \begin{center}
      \includegraphics[width=\columnwidth]{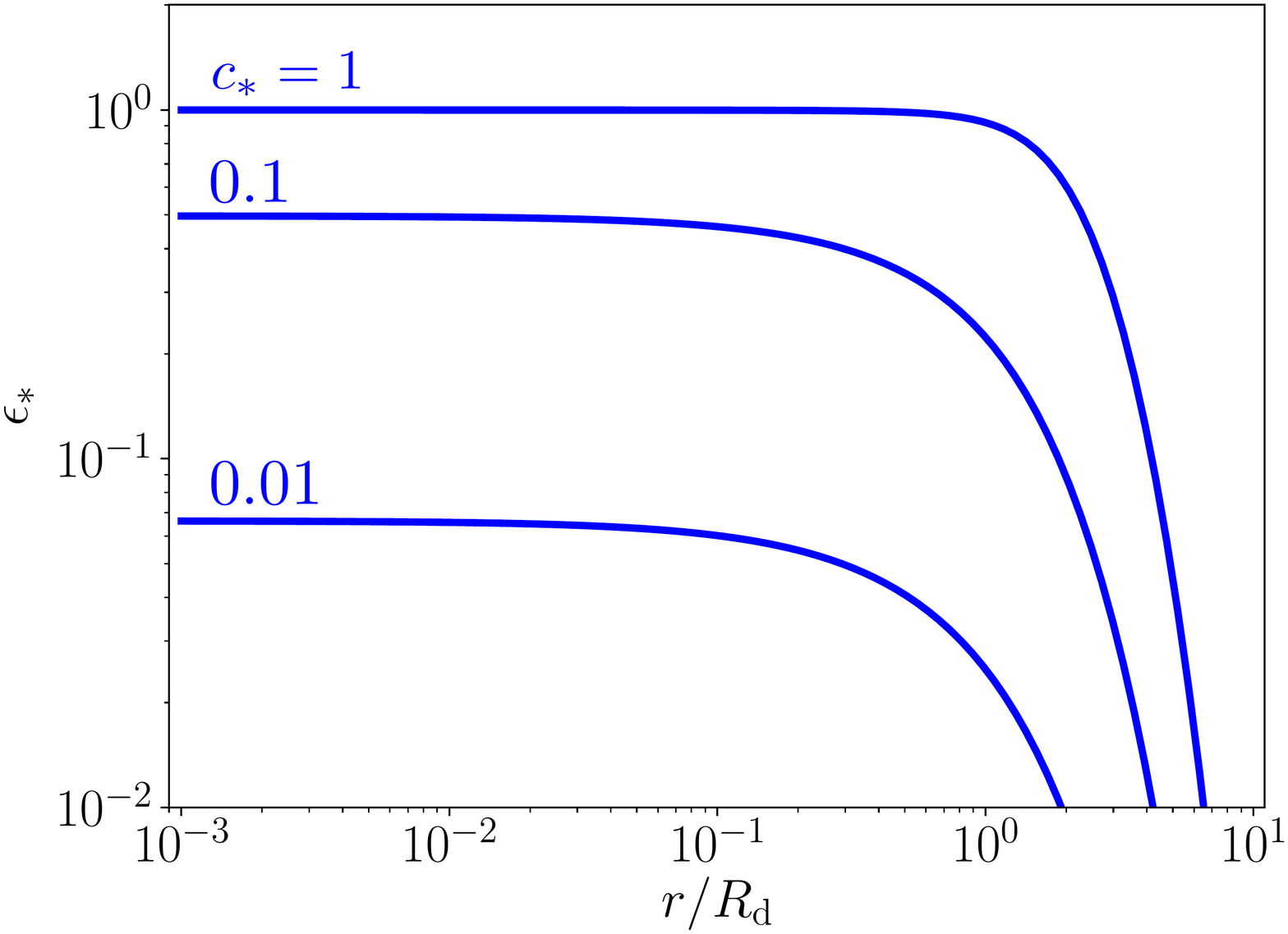}
    \end{center}
    \caption{ The radial distribution of SFE in a galactic disk with 
$M_{\rm h} = 10^{9} M_{\odot}$, $ 1 + z_{\rm vir} = 10$, $T_{\rm g} = 8000{\rm K}$, $f_{\rm d} = 0.5$ and $\lambda = 0.05$
    for three values of the star-formation rate parameter $c_{*} = 1, 0.1$, and $0.01$ from top to bottom.  \label{zu.0908.1}}
\end{figure}

As in Section \ref{eps_and_strformationrate}, we estimate the SFE $\epsilon_{*}$ by Equation \eqref{0723.4}.
Since the gas density does not decrease remarkably up to the scale height $z_{0}$ according to Equation \eqref{0814.1}, 
we use the density on the equatorial plane ($z=0$) in calculating $\epsilon_{*}$ in the disk.
The radial distribution of SFE is presented in Figure \ref{zu.0908.1} 
for three cases of $c_{*} = 1, 0.1$, and $0.01$. 
The SFE is almost constant inside the disk radius $R_{\rm d}$ and decreases rapidly outside.
In the case with $c_{*} = 1$, the SFE is as high as $\sim 1$ in the disk.
The stellar surface density is thus given by 
\begin{align}
	\Sigma_{*} &= \epsilon_{*} \Sigma_{\rm gas} & \nonumber \\
			 &=  2.2 \times 10^{2} \, M_{\odot}  {\rm pc^{-2}} \,   \left( \frac{f_{\rm d}}{0.5} \right) \left( \frac{\lambda}{0.05} \right)^{-2}  \nonumber \\
			 & \hspace{1cm} \times \left( \frac{1+z_{\rm vir}}{10} \right)^{2}  \left( \frac{\epsilon_{*}}{0.1} \right) \left( \frac{M_{\rm h}}{10^{9}M_{\odot}} \right)^{1/3} \exp \left( - \frac{r}{R_{\rm d}} \right).   \label{0816.5}
\end{align}

The luminosity and the ionizing photon emissivity per unit area, $\mathcal L$ and $\mathcal S$, 
can be calculated from the surface density of stars $\Sigma_{*}$ and luminosity/emissivity per unit stellar mass given in Table 1.
For example, in the case of $m_{\rm ch} = 10M_{\odot}$, they are 
\begin{align}
	\mathcal L = 1.5 \times 10^{6}  \,  L_{\odot} \, {\rm pc^{-2}} \, &   \left( \frac{f_{\rm d}}{0.5} \right)  \left( \frac{\lambda}{0.05} \right)^{-2} \left( \frac{1+z_{\rm vir}}{10} \right)^{2}  \nonumber \\
	 & \times \left( \frac{\epsilon_{*}}{0.1} \right) \left( \frac{M_{\rm h}}{10^{9}M_{\odot}} \right)^{1/3}  \exp \left( - \frac{r}{R_{\rm d}} \right),  \label{0816.6}    
\end{align}
and 
\begin{align}
	\mathcal S = 1.1 \times 10^{50}  \, {\rm s^{-1}} \,  {\rm pc^{-2}} \, &   \left( \frac{f_{\rm d}}{0.5} \right)  \left( \frac{\lambda}{0.05} \right)^{-2} \left( \frac{1+z_{\rm vir}}{10} \right)^{2} \nonumber \\
	& \times  \left( \frac{\epsilon_{*}}{0.1} \right) \left( \frac{M_{\rm h}}{10^{9}M_{\odot}} \right)^{1/3} \exp \left( - \frac{r}{R_{\rm d}} \right). \label{0816.7}
\end{align}

\subsection{Dust evacuation time from galactic disks}
We assume that stars are distributed on the mid-plane of galactic disks.
As in Section \ref{sec.det}, we derive the condition for dust evacuation by calculating the evacuation time $t_{\rm evac}$ 
at the scale height  $z = z_{0}$ (Eq. \ref{0814.2}).
We use the cylindrical coordinate $(r, \theta, z)$ where the $z$-axis passes through the center of the galaxy and perpendicular to the disks.
From the axisymmetry, no $\theta$-dependence appears.

Using the SFE $\epsilon_{*}$ (Eq. \ref{0723.4}), the luminosity and the ionizing photon emissivity 
per unit area in the disk are given as (Eqs. \ref{0816.6} and \ref{0816.7}),
\begin{eqnarray}
	\mathcal L = \mathcal L _{0} \left( 1- \exp \left[ - c_{*} \frac{t_{\rm OB}}{t_{\rm ff}} \exp \left( -r/R_{\rm d} \right) \right] \right) \exp \left( - r / R_{\rm d} \right), \label{0908.1}
\end{eqnarray} 
and
\begin{eqnarray}
	\mathcal S = \mathcal S _{0} \left( 1- \exp \left[ - c_{*} \frac{t_{\rm OB}}{t_{\rm ff}} \exp \left( -r/R_{\rm d} \right) \right] \right) \exp \left( - r / R_{\rm d} \right),  \label{0908.2}
\end{eqnarray} 
where
\begin{align}
	\mathcal L _{0} = 1.5 \times 10^{7} & \,  L_{\odot} \, {\rm pc^{-2}} \,   \left( \frac{f_{\rm d}}{0.5} \right)  \left( \frac{\lambda}{0.05} \right)^{-2} \left( \frac{1+z_{\rm vir}}{10} \right)^{2} \left( \frac{M_{\rm h}}{10^{9}M_{\odot}} \right)^{1/3},  \label{0908.3}    
\end{align}
and 
\begin{align}
	\mathcal S_{0} = 1.1 \times 10^{51} & \, {\rm s^{-1}} \,  {\rm pc^{-2}} \, \left( \frac{f_{\rm d}}{0.5} \right)  \left( \frac{\lambda}{0.05} \right)^{-2} \left( \frac{1+z_{\rm vir}}{10} \right)^{2}  \left( \frac{M_{\rm h}}{10^{9}M_{\odot}} \right)^{1/3}. \label{0908.3.2}
\end{align}

\begin{figure}
    \begin{center}
      \includegraphics[width=\columnwidth]{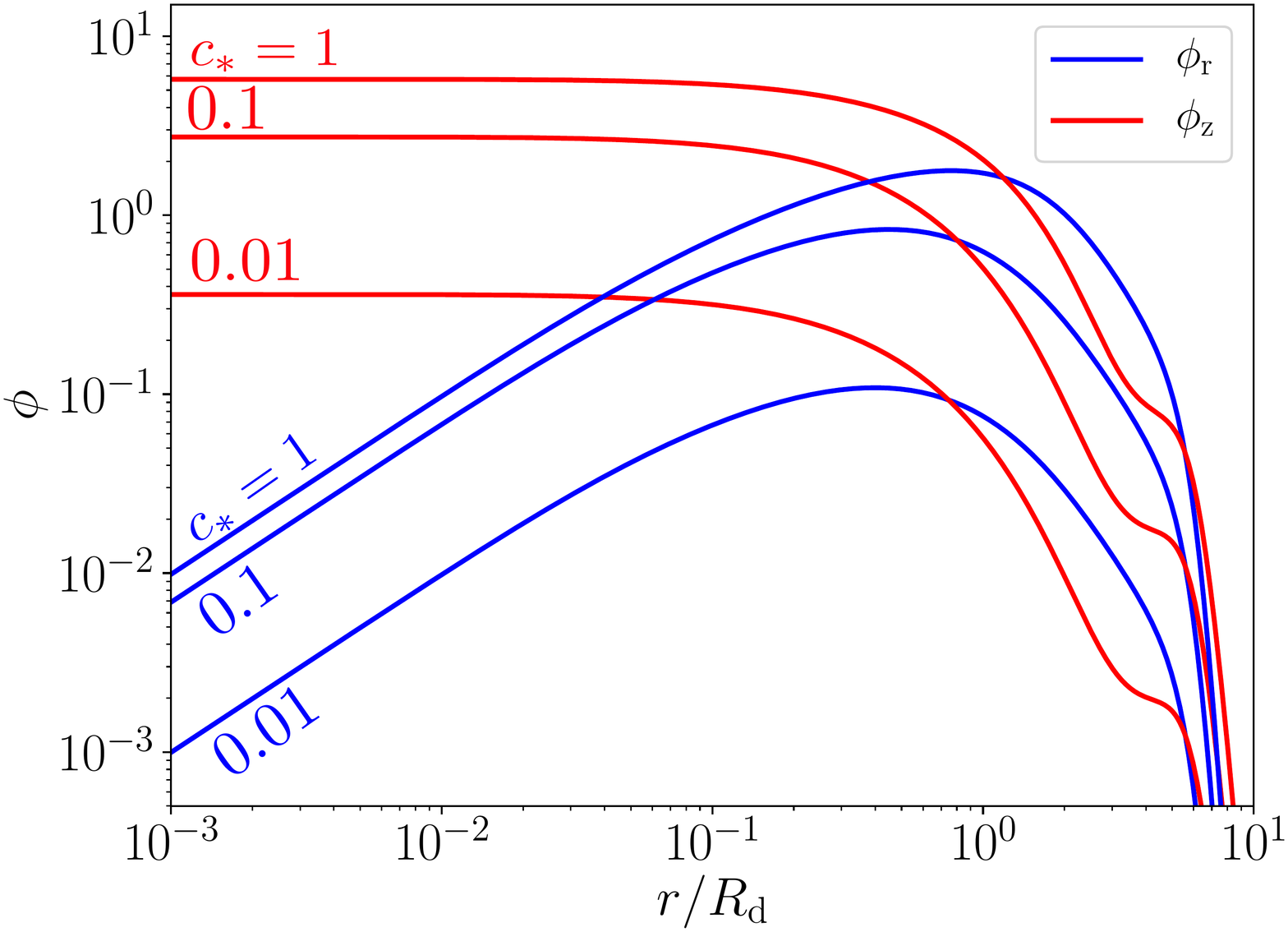}
        \caption{ The radial dependence of  $\phi_{r}$ (blue) and $\phi_{z}$ (red) in the case with 
    $\mathcal Z_{0} = 2.0 \times 10^{-2} \exp \left( \mathcal R_{0} \right)$. The different lines show the cases 
with $c_{*} = 1$, $0.1$, and $0.01$ from top to bottom. \label{zu0816.1}}
         \end{center}
\end{figure}

The $r$- and $z$-components of the radiation force on a dust grain with radius $a$ locating at $(r_{0},z_{0})$ are 
\begin{align}
        F_{{\rm rad}, r} &=  \frac{\pi a^{2} Q}{4 \pi c} \int ^{\infty}_{0} dr \int ^{2 \pi}_{0} r d \theta \frac{- (r \cos \theta - r_{0})}{\left[ r^{2} + r_{0}^{2} - 2 r r_{0} \cos \theta + z_{0}^{2} \right]^{3/2}} \mathcal L \nonumber \\
	& = \frac{\pi a^{2} Q \mathcal L_{0}}{4 \pi c} \phi_{r},  \label{0816.10}
\end{align}
and 
\begin{align}
	F_{{\rm rad}, z} &=  \frac{\pi a^{2} Q}{4 \pi c} \int ^{\infty}_{0} dr \int ^{2 \pi}_{0} r d \theta \frac{z_{0}}{\left[ r^{2} + r_{0}^{2} - 2 r r_{0} \cos \theta + z_{0}^{2} \right]^{3/2}} \mathcal L \nonumber \\
	& = \frac{\pi a^{2} Q \mathcal L_{0}}{4 \pi c} \phi_{z},  \label{0908.3.3}
\end{align}
respectively. 
With definition of $\mathcal R = r / R_{\rm d}$, $\mathcal R_{0} = r_{0} / R_{\rm d}, \mathcal Z_{0} = z_{0} / R_{\rm d}$ and 
$C = c_{*} t_{\rm OB} / t_{\rm ff}$, 
\begin{align}
	\phi_{r} = \int ^{\infty}_{0} d \mathcal R \int ^{2 \pi} _{0}  d \theta & \frac{(\mathcal R_{0} - \mathcal R \cos \theta) \mathcal R}{\left[ (\mathcal R^{2} + \mathcal R_{0}^{2} - 2 \mathcal R \mathcal R_{0} \cos \theta) + \mathcal Z_{0}^{2} \right]^{3/2}} \nonumber \\
	& \times  \left( 1 - \exp \left[ - C \exp \left( - \mathcal R \right) \right] \right) \exp \left( - \mathcal R \right), \label{0908.4}
\end{align}
and 
\begin{align}
	\phi_{z} = \int ^{\infty}_{0} d \mathcal R \int ^{2 \pi} _{0}  d \theta & \frac{\mathcal R \mathcal Z_{0}}{\left[ (\mathcal R^{2} + \mathcal R_{0}^{2} - 2 \mathcal R \mathcal R_{0} \cos \theta) + \mathcal Z_{0}^{2} \right]^{3/2}} \nonumber \\
	& \times  \left( 1 - \exp \left[ - C \exp \left( - \mathcal R \right) \right] \right) \exp \left( - \mathcal R \right). \label{0908.5}
\end{align}
The radial dependence of $\phi_{r}$ and $\phi_{z}$ is presented in Figure \ref{zu0816.1}.

\begin{figure}
    \begin{center}
      \includegraphics[width=\columnwidth]{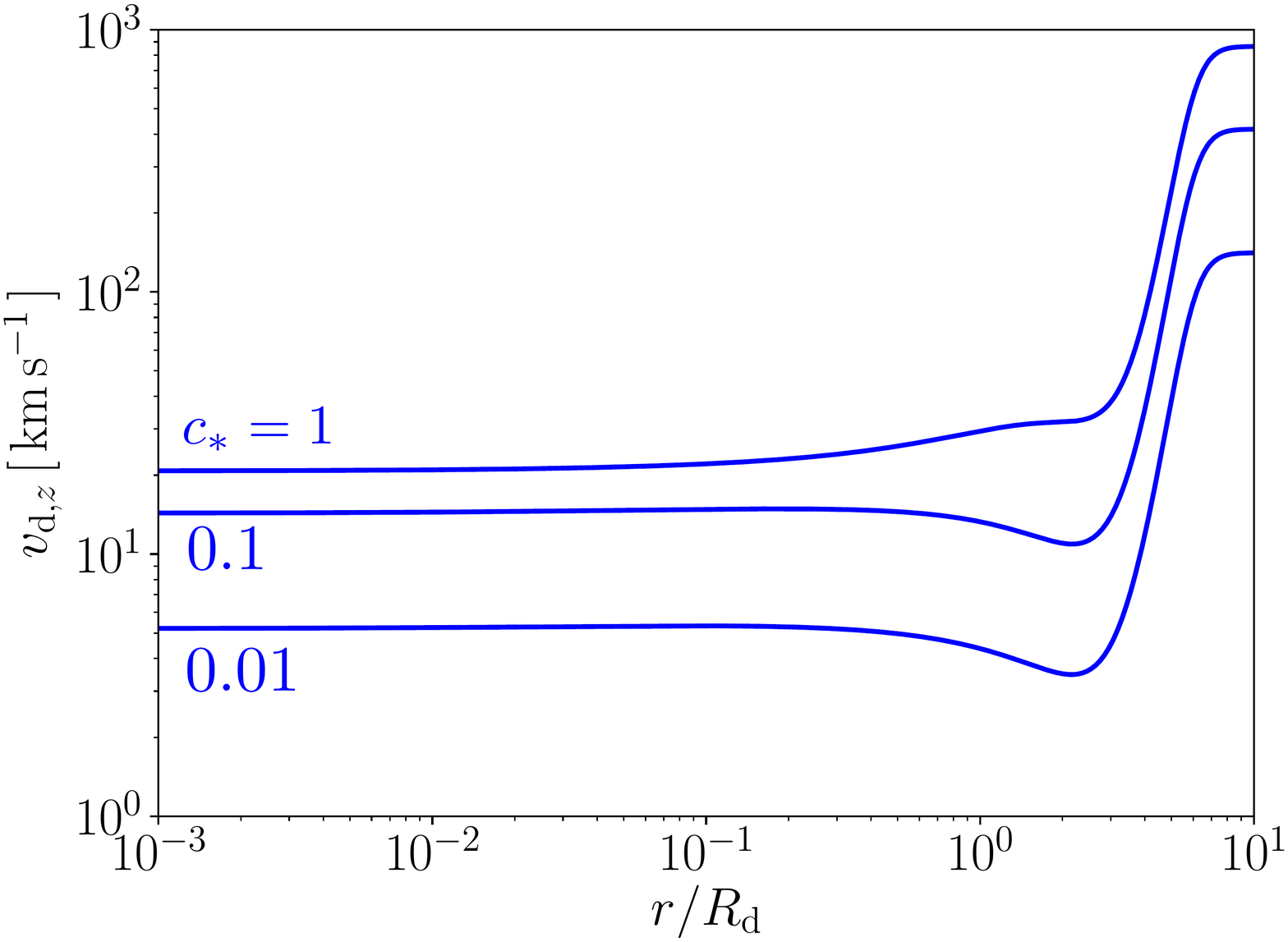}
        \caption{  The vertical component of the dust velocity $v_{{\rm d},z}$ at the scale height $z_{0}$ as a function of radius $r$  in the case with $f_{\rm d} = 0.5$, $\lambda = 0.05$, $1+z_{\rm vir} = 10$, $T_{\rm gas} = 8000~{\rm K}$, $Q = 1$, $\epsilon_{*} = 0.1$ and $M_{\rm h} = 10^{9}M_{\odot}$.
        The different lines show the case with $c_{*} = 1$, $0.1$ and $0.01$, from the top to the bottom.
        \label{zu0818.1}}
         \end{center}
\end{figure}

The dust terminal velocity is estimated by the balance between the
radiation force $F_{\rm rad} = \sqrt{\left(F_{\rm rad, r} \right)^{2} + \left(F_{\rm rad, z} \right)^{2}}$ and the collisional drag force 
$F_{\rm drag} = \pi a^{2} v_{\rm d}^{2} n_{\rm H} m_{\rm H}$.
Here, we consider the dust evacuation in the vertical direction.
The vertical component of terminal velocity $v_{{\rm d}, z}$ is 
\begin{eqnarray}
	v_{{\rm d}, z} = \sqrt{\frac{\mathcal L_{0} Q}{4 \pi c n_{\rm H} m_{\rm H}}} \phi_{\rm z} \left( \phi_{r}^{2} + \phi_{z}^{2} \right)^{-1/4}, \label{0816.13}
\end{eqnarray}
whose value at the center of the disk is 
\begin{align}
	v_{{\rm d}, z} (r=0) & = 1.4 \times 10^{6} {\rm cm \, s^{-1}}  \left( \frac{f_{\rm d}}{0.5} \right)^{-1/2} \left( \frac{\lambda}{0.05} \right) \left( \frac{1+z_{\rm vir}}{10} \right)^{-1} \nonumber \\
	&\times \left( \frac{\phi_{z}}{2.7} \right)^{1/2}  \left( \frac{T_{\rm gas}}{8000~{\rm K}} \right)^{1/2} \left( \frac{Q}{1} \right)^{1/2}  \left( \frac{M_{\rm h}}{10^{9}~M_{\odot}} \right)^{-1/6}.   \label{0816.14}
\end{align}
Figure \ref{zu0818.1} shows the radial dependence of the dust terminal velocity  $v_{\rm d, z}$ (Eq. \ref{0816.13}) at a disk scale height $(z=z_{0})$.
The terminal velocity is almost constant inside the disk radius $R_{\rm d}$,
and increases outward as the drag force becomes small with decreasing gas density.

\begin{figure}
    \begin{center}
      \includegraphics[width=\columnwidth]{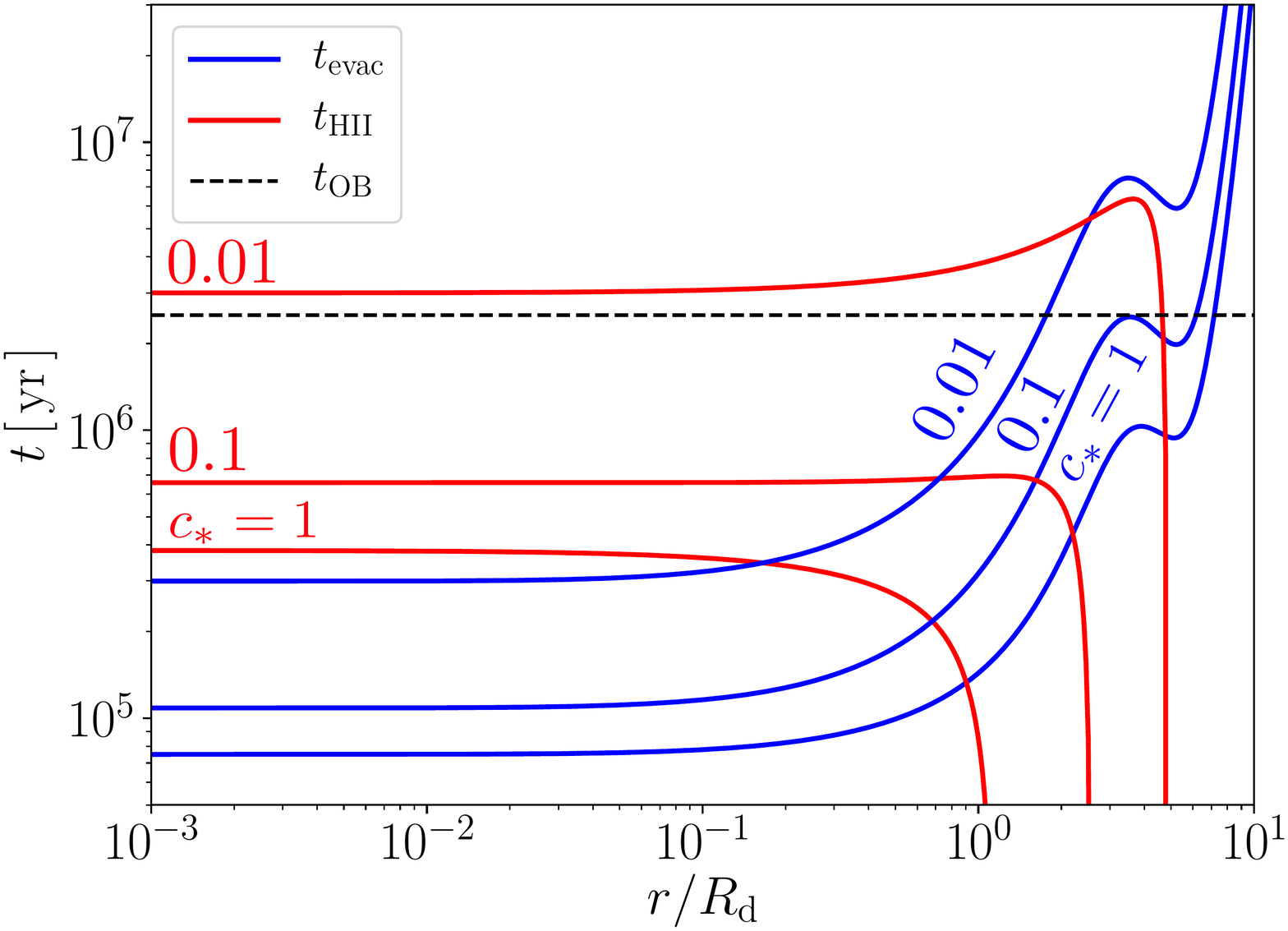}
        \caption{ Dust evacuation time  $t_{\rm evac}$, expansion time of H{\sc ii} regions $t_{\rm HII}$ and lifetime of OB stars $t_{\rm OB}$ 
at the disk scale height $z_{0}$ in the case with $f_{\rm d} = 0.5$, $\lambda = 0.05$, $1+z_{\rm vir} = 10$, $T= 8000{\rm K}$, $Q = 1$, $\epsilon_{*} = 0.1$ and $M_{\rm h} = 10^{9}M_{\odot}$.
         The cases with $c_{*} = 1$, $0.1$, and $0.01$ are shown as indicated for $t_{\rm evac}$ and $t_{\rm HII}$ \label{zu0818.2}.}
    \end{center}
\end{figure}

The dust evacuation time $t_{\rm evac}$ is given by
\begin{align}
	t_{\rm evac} &= \frac{z_{0}}{v_{{\rm d},z}} \nonumber \\
	& =   1.1 \times 10^{5} \, {\rm yr} \,  \left( \frac{f_{\rm d}}{0.5} \right)^{-1/2} \left( \frac{\lambda}{0.05} \right) \left( \frac{1+z_{\rm vir}}{10} \right)^{-1}  \left( \frac{\phi _{\rm z}}{2.7} \right)^{-1/2} \nonumber \\
	& \hspace{1.5cm} \times  \left( \frac{T_{\rm gas}}{8000 {\rm K}} \right)^{1/2} \left( \frac{Q}{1} \right)^{-1/2}   \left( \frac{M_{\rm h}}{10^{9}M_{\odot}} \right)^{-1/6}.  \label{0816.15}
 \end{align}
In Figure \ref{zu0818.2}, we show the radial dependence of dust evacuation time $t_{\rm evac}$. 
It is almost constant inside the disk radius $R_{\rm d}$ as expected from the behavior of $v_{{\rm d},z}$ and increases rapidly at $r>R_{\rm d}$ due to the increasing scale height (Eq. \ref{0816.3}).

\subsection{Vertical expansion of the H{\sc ii} region in the disk}\label{dust_less_gala}
We seek the condition for star formation from dust-free galactic disks by 
comparing the lifetime of OB stars, $t_{\rm OB}$, and the expansion time of the H{\sc ii} region, $t_{\rm HII}$, with the dust evacuation time, $t_{\rm evac}$, 
as in Section \ref{cloud_dest} and \ref{sec.dustless}.

We define the Str\"omgren height $z_{\rm S0}$ where the number of 
ionizing photons emitted from $z=0$ equals the integrated recombination rate of hydrogen atoms:
\begin{eqnarray}
	\int ^{z_{\rm S0}}_{-z_{\rm S0}} dz n_{\rm H}^{2} \alpha_{\rm B} = \mathcal S. \label{0818.1}
\end{eqnarray} 
The ionizing-photon emissivity per unit area $\mathcal S$ is given by Equation \eqref{0908.3}.
Neglecting the density variation with height and substituting Equation \eqref{0814.1}, \eqref{0816.2} and \eqref{0908.3} into \eqref{0818.1}, 
we obtain 
\begin{align}
	z_{\rm S0} =  1.9 \times & 10^{-2} \, {\rm pc}  \left( \frac{f_{\rm d}}{0.5} \right)^{-3}  \left( \frac{\lambda}{0.05} \right)^{6} \left( \frac{1+z_{\rm vir}}{10} \right)^{-6} \nonumber \\
	& \times  \left( \frac{T}{8000{\rm K}} \right)^{2}  \left( \frac{\epsilon_{*}}{0.1} \right)  \left( \frac{M_{\rm h}}{10^{9}M_{\odot}} \right)^{-1}  \exp \left( \frac{3 r}{R_{\rm d}} \right)  .\label{0818.2}
\end{align}
As in Equation \eqref{1.6.12}, the time for the H{\sc ii} region to expand up to $z=z_{\rm 0}$ is estimated as 
\begin{eqnarray}
	t_{\rm HII} = \frac{4}{7} \frac{z_{\rm S0}}{c_{\rm HII}} \left[ \left( \frac{z_{0}}{z_{\rm S0}} \right)^{7/4} -1 \right]. \label{0818.3}
\end{eqnarray}
In particular, at the center of the disk, 
\begin{align}
	t_{\rm HII} (r = 0) = 2.2 \times & 10^{6} \, {\rm yr}  \left( \frac{f_{\rm d}}{0.5} \right)^{1/2}  \left( \frac{\lambda}{0.05} \right)^{-1} \left( \frac{1+z_{\rm vir}}{10} \right) \nonumber \\
	& \times \left( \frac{T}{8000{\rm K}} \right)^{1/4}\left( \frac{\epsilon_{*}}{0.1} \right)^{-3/4}  \left( \frac{M_{\rm h}}{10^{9}M_{\odot}} \right)^{1/6} .
	\label{0818.4}
\end{align}
The expansion time of the H{\sc ii} region $t_{\rm HII}$ is shown in Figure \ref{zu0818.2} as a function of the radius.
At $r < R_{\rm d}$, $t_{\rm HII}$ is longer than $t_{\rm evac}$ because
the initial Str\"omgren height of the H{\sc ii} region $z_{\rm S0}$ (Eq. \ref{0818.2}) is much smaller than the disk scale height $z_{0}$.
As the radius exceeds $R_{\rm d}$, the scale height of the H{\sc ii} region $z_{\rm HII}$ rapidly increases, 
resulting in shorter $t_{\rm HII}$ than $t_{\rm evac}$.

\subsection{Condition for dust evacuation from galactic disks}

For the dust evacuation (Eq. \ref{0907.2}), the evacuation time $t_{\rm evac}$ must be shorter than lifetime of the cloud, $t_{\rm cl} = {\rm min} (t_{\rm OB}, t_{\rm HII})$.
For example, in Figure \ref{zu0818.2} for the halo shown ($M_{\rm h} = 10^{9}~M_{\odot}$, $1+z_{\rm vir} = 10$, $T_{\rm gas} = 8000~{\rm K}$, $\lambda = 0.05$ and $f_{\rm d} = 0.5$), the dust evacuation occurs inside the galactic disk $r < R_{\rm d}$ and the dust-free gas is formed in all the three cases.

Additionally, there is a condition on metallicity of the gas. 
As explained in Section \ref{sec.det}, the direct light from stars is absorbed and re-emitted 
as thermal emission of dust grains in optically thick disks,
and radiation force on dust grains becomes about two orders of magnitude smaller.
Therefore, the optical depth to the direct light along the vertical direction must be less than unity for the dust evacuation:
\begin{align}
	\tau &= \int ^{\infty}_{-\infty} \rho \kappa dz \nonumber \\
	       &= 1.6 \times 10^{2}  \left( \frac{Z}{Z_{\odot}} \right) \left( \frac{f_{\rm d}}{0.5} \right) \left( \frac{\lambda}{0.05} \right)^{-2} \left( \frac{1+z_{\rm vir}}{10} \right)^{2} \nonumber \\
	       & \hspace{1.5cm} \times  \left( \frac{\kappa}{350 {\rm cm^{2} g^{-1}}} \right) \left( \frac{M_{\rm h}}{10^{9}M_{\odot}} \right)^{1/3} \exp \left( \frac{-2r}{R_{\rm d}} \right) < 1. \label{0818.5}
\end{align}
This requires low metallicity environments of
\begin{align}
 Z < 6.1 \times 10^{-3}Z_{\odot}  \left( \frac{f_{\rm d}}{0.5} \right)^{-1} & \left( \frac{\lambda}{0.05} \right)^{2} \left( \frac{1+z_{\rm vir}}{10} \right)^{-2} \nonumber \\
  & \times \left( \frac{\kappa}{350 {\rm cm^{2} g^{-1}}} \right)^{-1} \left( \frac{M_{\rm h}}{10^{9}M_{\odot}} \right)^{-1/3}. \label{0818.6}
\end{align}
 
\begin{figure}
    \begin{center}
      \includegraphics[width=\columnwidth]{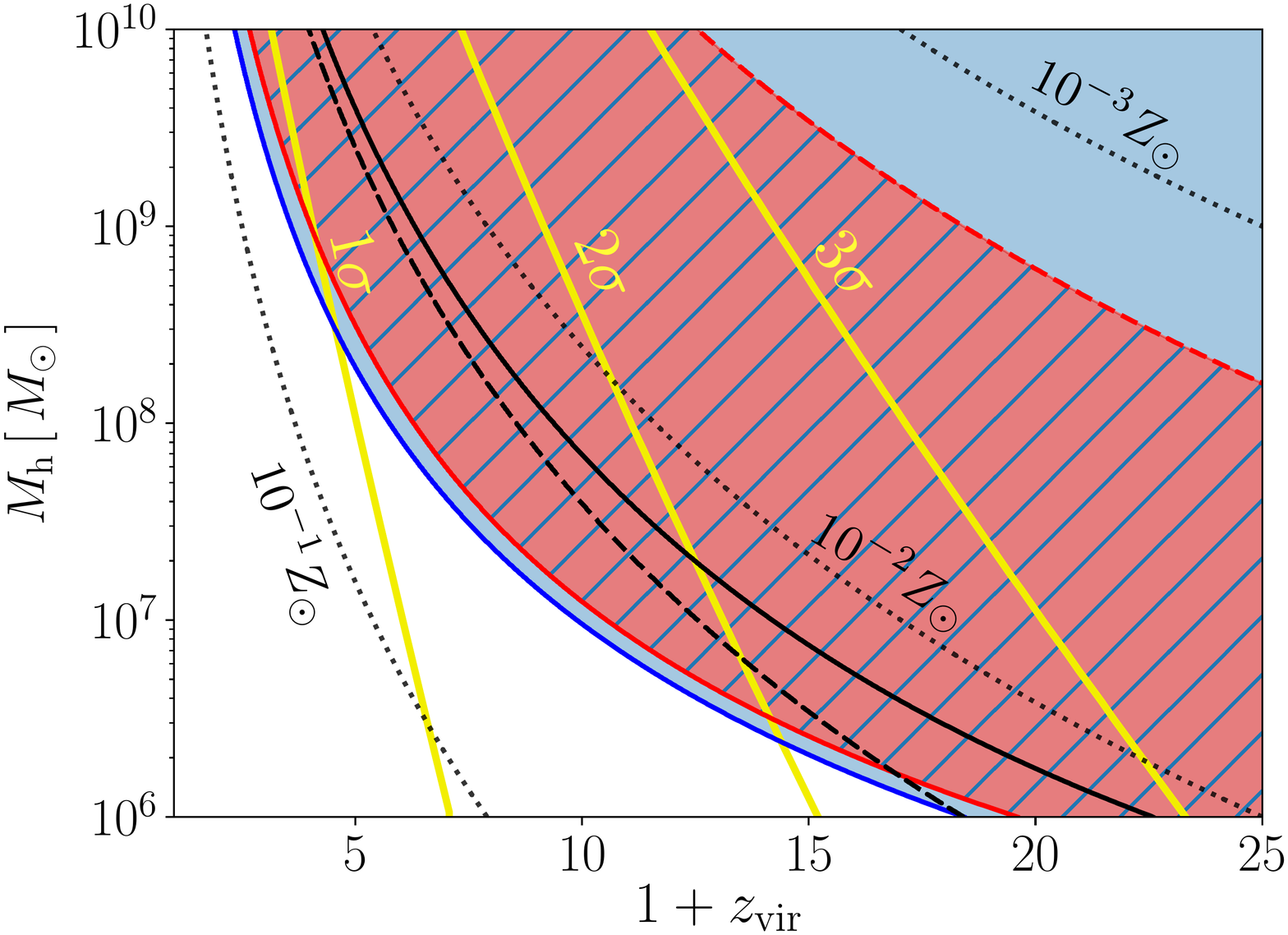}
          \caption{ Conditions for the dust evacuation from galactic disks.
          The thick solid lines correspond to the hallo mass $M_{\rm h}$ satisfying the condition $t_{\rm evac} = t_{\rm cl}$.
          The thick dashed lines show the galaxies where the SFE becomes $\epsilon_{*} = 0.9$.
          Black, red and blue colors correspond to the cases with $c_{*} = 1$, $0.1$ and $0.01$, respectively.
          The color-shaded areas between them represent the regions that satisfy both the conditions $t_{\rm evac} < t_{\rm cl}$ and $\epsilon_{*} < 0.9$.
          Namely, those conditions are satisfied in the upper side of the blue-solid line (both in the blue and red shaded regions) for $c_{*} = 0.1$, and between the red-solid and red-dashed lines (in the red-shaded region) for $c_{*}=0.01$, respectively.
          In the case with $c_{*} = 1$, there is no such region. 
 	 The galactic disks become optically thin below the thin dotted lines for the indicated metallicity.
          The yellow-solid lines show the halo mass and redshift when $1$, $2$ and $3~\sigma$ fluctuations are virialized 
           \citep{Barkana2001,Eisenstein1998,Eisenstein1999,Planck2016}.
    \label{zu0908_2}}
      \end{center}
\end{figure}

Next we derive the constraint on the halo mass $M_{\rm h}$ and redshift $1+z_{\rm vir}$
under the condition of the spin parameter $\lambda = 0.05$ and 
the baryon mass fraction taken into the disks $f_{\rm d} = 0.5$.
Figure \ref{zu0908_2} shows the range of halo parameters for the dust evacuation from the disks.
Note that since the timescale $t_{\rm evac}$ and $t_{\rm cl}$ hardly change inside the disk $r < R_{\rm d}$, 
it suffices to consider the condition $t_{\rm evac}<t_{\rm cl}$ at the center ($r=0$). 
In Figure \ref{zu0908_2}, black, red and blue colors correspond to the cases with the SFR parameter $c_{*} = 1, 0.1$, and $0.01$, respectively.
The condition $t_{\rm evac}<t_{\rm cl}$ is satisfied above the solid line, but  
above the dashed line the SFE is already $>0.9$ at $t_{\rm evac}$.
Thus, the color-shaded regions between the two lines represents the parameter space where the dust evacuation occurs 
and gases still remain at this time, resulting in star formation from the dust-free gas.
Also the thin-dotted lines show the contour of the maximum metallicity allowed (Eq. \ref{0818.6}) for a dust evacuation 
for a given parameter set ($M_{\rm h}, z_{\rm vir}$).
Additionally, the yellow-solid lines show the relation between the halo mass and the formation redshift for the $1$, $2$ and $3 \sigma$ cosmological density fluctuations. 
Here, we use the cosmological parameters of \citet{Planck2016} and \citet{Eisenstein1998,Eisenstein1999}.

For higher halo mass and formation redshift, the dust evacuation time $t_{\rm evac}$ is shorter (Eq. \ref{0816.15}) and the H{\sc ii} region expansion time $t_{\rm HII}$ is longer (Eq. \ref{0818.4}).
Thus the minimum halo mass for the dust evacuation increases with decreasing redshift.
As an example, for a halo with mass $M_{\rm h} = 10^{9}M_{\odot}$, 
the condition for dust evacuation is satisfied if formed at $z_{\rm vir} \gtrsim 3$
with metallicity $Z \lesssim 10^{-1} - 10^{-2}~Z_{\odot}$
with little dependence on $c_{\ast}$.
With lower halo mass, formation redshift must be earlier, 
e.g., $ z_{\rm vir} \gtrsim 9$ with $M_{\rm h}=10^{7}M_{\odot}$.
As shown in Figure \ref{zu0908_2}, the lowest halo mass for the dust evacuation is equivalent to the $2 \sigma$ ($1 \sigma$) fluctuation at $z_{\rm vir} \simeq 15$ ($\simeq 5$, respectively).
With larger value of $c_{*}$, the H{\sc ii} region expansion time, $t_{\rm HII}$, becomes shorter.
In addition, since the galactic disk must be optically thin, 
the dust evacuation only occurs at metallicity lower than $\sim 10^{-2}Z_{\sun}$, 
depending somewhat on the halo mass and formation redshift.

\section{Caveats} \label{suppression}
In Sections \ref{dust_evac1} and \ref{dustevac2}, 
we have modelled the dust evacuation under the assumption that the ionization degree 
is lower than $\sim 10^{-5}$ so that the Coulomb drag is inefficient. 
The cosmic rays (CRs) or X-rays from nearby star-forming galaxies, however, can boost 
the ionization degree in the interstellar medium.
In addition, although we assumed the efficiency factor of dust photo-absorption $Q=1$
in calculating the evacuation time, it becomes smaller for grains with size $ \la 0.01 {\rm \mu m}$ 
and the radiation force on dust becomes weaker. 
Also, we have not considered magnetic fields, which prevents the perpendicular motion of charged grains. 
In the following, we discuss those effects on the dust evacuation.

\subsection{Possible enhancement of ionization degree} \label{sec4.1}
As discussed in Section \ref{sec.det}, the Coulomb drag suppresses the motion of dust grains and makes 
the evacuation time longer than the lifetime of massive stars for ionization degree $x_{\rm e} > 10^{-5}$.
The increase of ionization degree is likely caused by the cosmic rays, X-rays 
from nearby galaxies, or mixing with ionized gas from H{\sc ii} regions or stellar winds.
In the following, we estimate their impact on the ionization degree.

\subsubsection{Cosmic rays}
We here follow \citet{Stacy2007} for estimating the ionization degree by CRs, 
which were supposed to be produced in SN remnants in a different nearby galaxy.
Since we are considering the dust evacuation before the first SN explodes, 
we assume the CRs and X-rays are coming from other nearby galaxies.
The ionization rate $\xi_{\rm CR}$ is related with the energy density of CRs $U_{\rm CR}$ as:
\begin{eqnarray}
  \xi_{\rm CR} = 1.4 \times 10^{-17} {\rm s^{-1}} \left( \frac{U_{\rm CR}}{10^{-15} \, {\rm erg \, cm^{-3}}} \right), \label{0121.1}
\end{eqnarray}
where the energy distribution is assumed to be the same power law as in the Milky Way with the slope $-2.65$ \citep{Blumer2009,Draine2011a},
ranging from $10^{6} \, {\rm eV}$ to $10^{15} \, {\rm eV}$.
Assuming an efficiency 0.1 of SN energy $10^{51}{\rm erg}$ is used in the CR acceleration, 
we adopt the total CR energy generated in a SN remnant $E_{\rm CR} = 10^{50} \, {\rm erg}$.
Using the mass fraction of massive stars $f_{\rm OB}=0.74$ 
and their average mass $\bar m_{\rm OB}=29M_{\sun}$ for the IMF of Equation \eqref{3.1.1}, 
the SN rate is 
\begin{align}
  \dot N_{\rm SN} &= \frac{f_{\rm OB} {\rm SFR}}{\bar m_{\rm OB}} \nonumber \\ 
                  &= 2.6 \times 10^{-2} \, {\rm yr^{-1}} \left( \frac{{\rm SFR}}{1 \, M_{\odot} \, {\rm yr^{-1}}} \right). \label{0122.2}
\end{align}
At the distance $d$ from the CR source galaxy, 
the CR energy density is given by 
\begin{align}
  U_{\rm CR} &= \frac{\dot N_{\rm SN } E_{\rm CR}}{4 \pi d^2 v_{\rm CR}} \nonumber \\ 
             &= 3.4 \times 10^{-15} \, {\rm erg \, cm^{-3}} \left( \frac{d}{10 \, {\rm kpc}} \right)^{-2} \left( \frac{{\rm SFR}}{1 \, M_{\odot} {\rm yr^{-1}}} \right), \label{0122.3}
\end{align}
where $v_{\rm CR} = 8.8 \times 10^{-2} c$ is the average velocity of CRs \citep{Stacy2007}.
By substituting Equation  \eqref{0122.3} into Equation \eqref{0121.1},
we obtain the CR ionization rate:
\begin{eqnarray}
  \xi_{\rm CR} = 4.8 \times 10^{-17} \, {\rm s^{-1}} \, \left( \frac{d}{10 \, {\rm kpc}} \right)^{-2} \left( \frac{{\rm SFR}}{1 \, M_{\odot} {\rm yr^{-1}}} \right). \label{0122.4}
\end{eqnarray}
Using this rate in Equation (\ref{0802.2}), 
the ionization degree is given by
\begin{align}
  x_{\rm e} = & 3.5 \times 10^{-5}  \left( \frac{R_{\rm cl}}{10 ~ {\rm pc}} \right)^{3/2} \nonumber \\
                                 & \times \left( \frac{M_{\rm cl}}{10^{6} ~ M_{\odot}} \right)^{-1/2} 
    \left( \frac{d}{10 ~ {\rm kpc}} \right)^{-1} \left( \frac{{\rm SFR}}{1 ~ M_{\odot} {\rm yr^{-1}}} \right)^{1/2}. \label{0122.5}
\end{align}
This means that if the source galaxy with ${\rm SFR} \gtrsim 1~M_{\odot} {\rm yr^{-1}}$ is within 40 kpc, the CRs raise the ionization degree above the threshold value $10^{-5}$ for the Coulomb drag, resulting in the suppression of the dust evacuation.

In the above discussion, we assumed that CRs stream away freely from the source galaxy.
Magnetic fields, if present with sufficient strength, confine CRs in the source galaxy.
Given that SN remnants are inside the galaxy, the CR intensity is enhanced not only by their proximity but also by this magnetic effect.
In this case, the ionization degree becomes much higher compared with Equation \eqref{0122.5}, and the dust evacuation is strongly inhibited.
In contrast, if the CR sources are outside the galaxy under consideration, the ionization rate would be reduced because the magnetic field protects the galaxy from the CR incidence.

\subsubsection{X-rays}
Massive binary stars can evolve to X-ray binaries. 
We evaluate the ionization by such X-ray sources based on the model of \citet{Glover2003}.
We again assume the sources are in a nearby galaxy at distance $d$ from the cloud under consideration. 
The X-ray luminosity is related to the star formation rate as 
\begin{eqnarray}
  L_{\rm X} = 6.0 \times 10^{38} ~ {\rm erg \, s^{-1}} \left( \frac{{\rm SFR}}{1~M_{\odot} {\rm yr^{-1}}} \right). \label{0122.6}
\end{eqnarray}
For the power-law spectrum with the slope $- 1.5$, 
the X-ray intensity is given by 
\begin{align}
  J_{\rm X} = 2.3 \times 10^{-25} & ~ {\rm erg \, s^{-1} \,  cm^{-2} \, sr^{-1} \, Hz^{-1}} \nonumber \\
    & \times \left(\frac{\nu}{\nu_{0}} \right)^{-1.5} \left( \frac{d}{10 \, {\rm kpc}} \right)^{-2}  \left( \frac{{\rm SFR}}{1 ~ M_{\odot} {\rm yr^{-1}}} \right), \label{0122.7}
\end{align}
where $h \nu_{\rm 0} = 1~{\rm eV}$.
The primary ionization rate of X-rays is thus given by 
\begin{align}
  \xi_{\rm X,p} &= \int \frac{4 \pi J_{\rm X}(\nu)}{h \nu} \sigma_{\rm H} (\nu) d \nu \nonumber \\ 
                &= 6.8 \times 10^{-23} ~ {\rm s^{-1}} \left( \frac{d}{10~{\rm kpc}} \right)^{-2} \left( \frac{{\rm SFR}}{1~M_{\odot}{\rm yr^{-1}}} \right), \label{0122.8}
\end{align}
where $\sigma_{\rm H}(\nu)$ is the cross section of hydrogen and we have set the energy range from $2~{\rm keV}$ to $10 ~{\rm keV}$.
Secondary ionization is dominant in the case of X-ray ionization.
Using the secondary ionization rate $\phi^{\rm H}$ by \citet{Wolfire1995}, we obtain the X-ray ionization rate as 
\begin{align}
  \xi_{\rm X} &= \xi_{\rm X,p} (1+\phi^{\rm H}) \nonumber \\ 
              &= 8.8 \times 10^{-21} ~{\rm s^{-1}} \left( \frac{d}{10~{\rm kpc}} \right)^{-2} \left( \frac{\rm SFR}{1~M_{\odot}{\rm yr^{-1}}} \right). \label{0122.9}
\end{align}
Comparing with the CR ionization (Eq. \ref{0122.4}), 
X-ray ionization rate is more than two orders of magnitude lower and 
has little impact on the dust evacuation.

\subsubsection{Mixing with ionized medium}
Near the boundary with ionized gases such as the H{\sc ii} regions or ionized stellar winds, 
mixing with neutral and ionized medium may occur due, e.g., to hydrodynamical instabilities. 
Although this raises the ionization degree in the neutral gas temporarily,  
it decreases in the recombination time, which 
can be estimated for the ionization degree $x_{\rm e} = 10^{-5}$ as
\begin{align}
  t_{\rm rec} &= \left( \alpha_{\rm B} n_{\rm H} x_{\rm e} \right)^{-1} \nonumber \\
              &= 4.5 \times 10^{4} ~{\rm yr} \left( \frac{n_{\rm H}}{10^{4} ~{\rm cm^{-3}}} \right)^{-1} \left( \frac{x_{\rm e}}{10^{-5}}\right)^{-1}. \label{0122.10} 
\end{align}
Since the recombination time is much shorter than the dust evacuation time 
$t_{\rm evac}$ (Eq. \ref{1.6.4}, \ref{0816.15}), mixing with the ionized gas does not affect the dust evacuation 
unless the ionized gas is constantly supplied by some mechanism in a timescale shorter than $\sim t_{\rm rec}$.

\subsection{Small dust grains} \label{sec4.2}
\begin{figure}
    \begin{center}
      \includegraphics[width=\columnwidth]{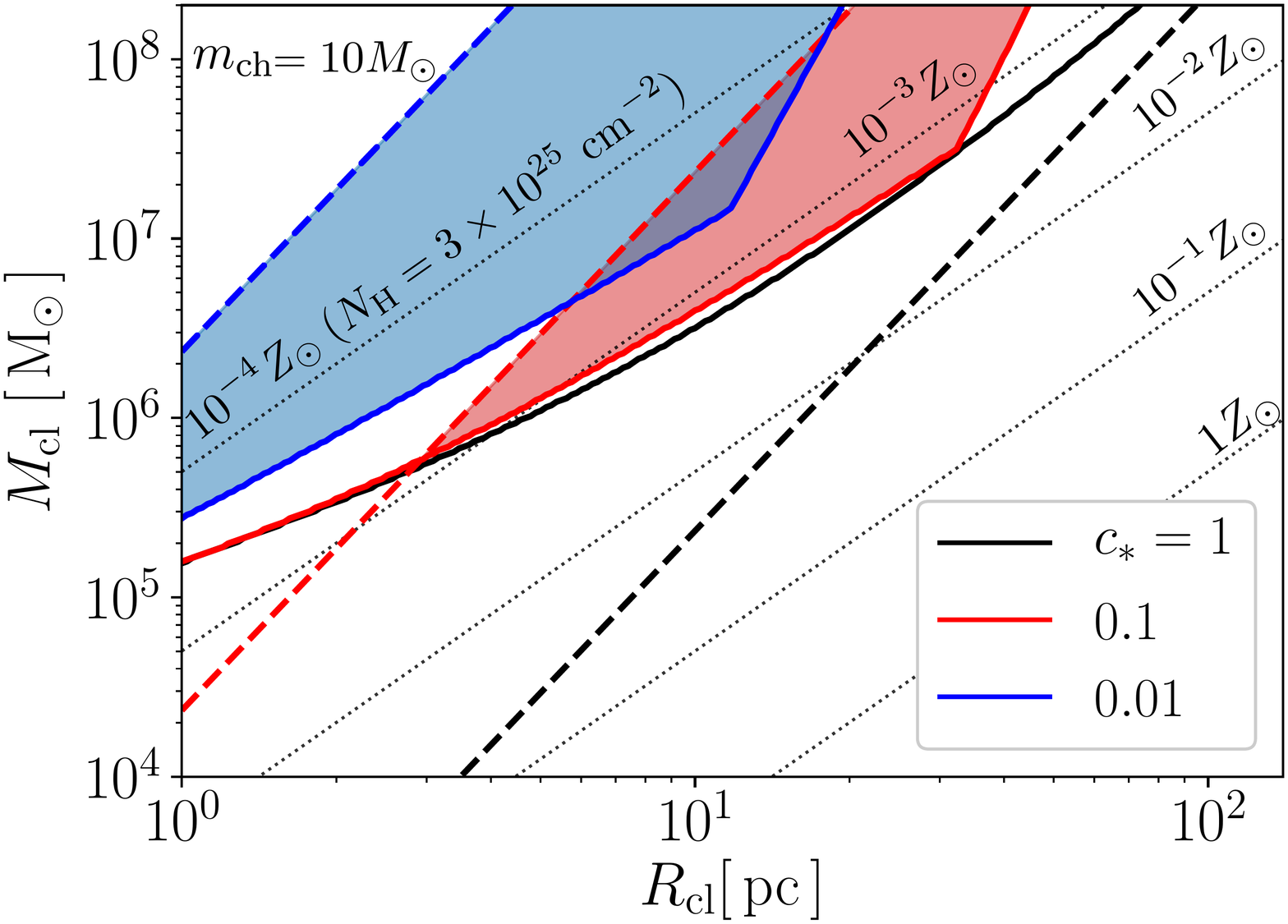}
          \caption{Same as Figure \ref{zu0807.3}, but for the case of the smaller efficiency factor of $Q = 0.3$.  \label{zu0123.1}}
      \end{center}
\end{figure}
In previous sections, we assumed that the dust absorption efficiency factor $Q=1$, corresponding to grains 
larger than $\sim 0.01~\mu{\rm m}$. 
If small grains are dominant, however, the efficiency factor decreases and the dust evacuation time becomes longer as $Q^{-1/2}$ 
(Eqs. \ref{1.6.4} and \ref{0816.15}). 

Dust grains in the early Universe (at $z>6$) are mainly produced in SN events because intermediate-mass stars ($<8~M_{\odot}$) take more than $1~{\rm Gyr}$ to become AGB stars which produce a large quantity of dust grains \citep{Morgan2003,Dwek2007}. 
Theoretical calculations indicate that small grains $a< 0.01~\mu{\rm m}$
have a flat ''top-heavy'' size distribution in the early Universe \citep{Todini2001,Schneider2012}, 
unlike the so-called MRN distribution \citep{Mathis1977} in the 
local interstellar medium, which has a ``bottom-heavy''power-law with the slope $\alpha = -3.5$ in the range $a \sim 0.01 - 1.0~\mu{\rm m}$. 
In addition, small grains $\la 0.05~\mu{\rm m}$ are subsequently destroyed by the passage of the SN reverse shock 
in their formation site \citep{Nozawa2007,Bianchi2007}.
Since the efficiency factor $Q \sim 1$ (Figure \ref{zu0807.6}) for grains larger than $0.01~\mu{\rm m}$, 
our assumption $Q=1$ would be reasonable for dust in the early universe.
Considering uncertainties in the dust size distribution, however, 
we also calculate the case of small grains with $a=5\times10^{-3}~\mu{\rm m}$, which have $Q \simeq 0.3$.
Figure \ref{zu0123.1} presents the condition for formation of dust-free star-forming clouds.
Due to smaller efficiency factor, the parameter range for the dust-free cloud formation becomes smaller 
compared to the fiducial case (Figure \ref{zu0807.3}).
Only massive compact clouds satisfy the dust evacuation condition 
if the typical dust size is smaller than $\sim 0.01 \mu{\rm m}$.
Further studies are needed about the size distribution of grains in the early Universe.

\subsection{Magnetic field} \label{sec4.3}
Dust grains have positive electric charge due to photoelectric effect so that 
their motion is influenced by the presence of magnetic field. 
In fact, the gyro-radius of a grain
\begin{align}
  r_{\rm gyro} &= \frac{m_{\rm d} v_{\rm d} c}{q B} \nonumber \\
               &= 9.4 \times 10^{-7} ~{\rm pc} \left( \frac{q}{600~e} \right)^{-1} \left( \frac{B}{100~\mu{\rm G}} \right)^{-1} \nonumber \\
               & \hspace{2cm} \times \left( \frac{\rho_{\rm d}}{3~{\rm g \, cm^{-3}}} \right) \left( \frac{a_{\rm d}}{0.1~\mu{\rm m}} \right)^{3} \left( \frac{v_{\rm d}}{7 ~ {\rm km \, s^{-1}}} \right), \label{0124.1}
\end{align}
is much smaller than the star-forming regions
if the magnetic field is as strong as in local molecular clouds $\sim 100 ~\mu{\rm G}$ \citep{Crutcher1991}.

Below, we evaluate the dust drift velocity taking the effect of magnetic field into account \citep{Draine2011}.
For simplicity, we assume that the magnetic field is parallel to the $z$-axis, $\bm{B} = (0,0,B)$, 
and the radiation force direction is at angle $\theta$ to the field orientation, $\bm{F}_{\rm rad}=(F_{\rm rad} \sin \theta,0,F_{\rm rad} \cos \theta)$.
Dust grains have steady time-averaged velocity $\bar{\bm{v}}$ with rotational motion around the magnetic field,   
with $x$, $y$, and $z$-components, respectively, 
\begin{eqnarray}
	\bar{v}_{\rm x} &=& \frac{1}{1 + ( \omega t_{\rm drag} )^{2}} \frac{F_{\rm rad} t_{\rm drag}}{m_{\rm d}} \sin \theta, \label{0131.1} \\
	\bar{v}_{\rm y} &=& - \frac{(\omega t_{\rm drag})}{1 + ( \omega t_{\rm drag} )^{2}} \frac{F_{\rm rad} t_{\rm drag}}{m_{\rm d}} \sin \theta,\label{0131.2} \\
	\bar{v}_{\rm z} &=& \frac{F_{\rm rad} t_{\rm drag}}{m_{\rm d}} \cos \theta.  \label{0131.3}
\end{eqnarray}
The drift velocity becomes 
\begin{eqnarray}
	v_{\rm d} = \sqrt{\frac{1 + \cos ^{2} \theta (\omega t_{\rm drag})^{2}}{1 + (\omega t_{\rm drag})^{2}}} \frac{F_{\rm rad} t_{\rm drag}}{m_{\rm d}}, \label{0131.4}
\end{eqnarray}
where $\omega=qB/m_{\rm d}c$ is the gyro-frequency and $t_{\rm drag}$ is the drag time (Eq. \ref{0807.1}).
From Equation  \eqref{0131.4}, we can see that the magnetic field greatly hinders the grain motion
when
\begin{align}
  \omega t_{\rm drag} &= 0.26 \left( \frac{q}{600~e} \right) \left( \frac{B}{0.1~\mu{\rm G}} \right) \left( \frac{a_{\rm d}}{0.1~\mu{\rm m}} \right)^{-2} \nonumber \\
                      & \hspace{2cm} \times \left( \frac{v_{\rm d}}{7 ~ {\rm km \, s^{-1}}} \right)^{-1} \left( \frac{n_{\rm H}}{10^{4}~{\rm cm^{-3}}} \right)^{-1} \label{0124.2}
\end{align}
is much larger than unity.
In particular, the dust velocity is reduced by a factor of $1/\sqrt{1 + (\omega t_{\rm drag})^{2}}$  
from the value without a magnetic field when the direction of the radiation force is vertical to the field.
Magnetic fields in first galaxies are usually expected to be much weaker than in the Milky Way \citep{Koh2016}.
As seen in Equation \eqref{0124.2}, as long as the field is weaker than $\sim 0.1 ~\mu {\rm G}$, $\omega t_{\rm drag} \leq 1$ and  
the dust evacuation is not suppressed. 
Magnetic fields, however, can be potentially amplified by turbulent small-scale dynamo to equipartition, which corresponds to $\sim 100 ~ {\mu G}$ \citep{Schober2012}.
In this case, the dust drift in the vertical direction is strongly suppressed but dust grains can still drift in the direction along magnetic fields.
Thus, the suppression effect by magnetic fields would remain at the factor of a few level.
For further discussion on this problem, detailed numerical calculation is awaited.

\section{Summary and discussion}\label{matome}
We have investigated whether the dust grains are evacuated from star-forming regions 
by stellar radiation feedback so that star formation from the dust-free gas ensues 
in some environments in the early universe. 
We have obtained the condition for the dust evacuation by comparing the dust evacuation time, $t_{\rm evac}$, with the smaller of 
the H{\sc ii} region expansion time, $t_{\rm HII}$, and the OB star lifetime $t_{\rm OB}$.

Our findings are summarized as follows:

\begin{itemize}
\item[(1)] 

As star-forming clouds become compact, the evacuation time decreases.
Therefore, the dust evacuation can occur for compact star-forming clouds whose column density is $N_{\rm H} \simeq 10^{24} \sim 10^{26}~{\rm cm^{-2}}$, corresponding to the radius of $1 - 10~{\rm pc}$ for the mass $M_{\rm cl} = 10^{6}M_{\odot}$.
The radiation force on dust grains is reduced significantly if the clouds are optically thick to the radiation from stars.
This imposes the condition that metallicity should be less than $\sim 10^{-2}~Z_{\odot}$.

\item[(2)] 
 The dust evacuation from galactic disks occurs more easily in more massive halos and at earlier formation redshift.
For example, halos of $\sim 10^{9}~M_{\odot}$ ($\sim 10^{7}~M_{\odot}$) formed at $z \sim 3$ ($z \sim 9$, respectively) can induce the dust evacuation. 
We find that $t_{\rm evac}$ and $t_{\rm HII}$ are almost constant inside the galactic disks. 
Therefore the dust evacuation occurs in the entire galactic disks once the condition for the dust evacuation is satisfied at the center of disks.

\end{itemize}

We expect that the dust-to-gas mass ratio is reduced remarkably or even becomes zero due to the dust evacuation. 
On the other hand, the dust depletion factor $f_{\rm dep}$ is  $\sim 0.5$ in the solar neightborhood \citep{Pollack1994}, and it is theoretically suggested to be smaller in very low-metallicity environments \citep{Asano2013,Remy-Ruyer2014}.
Thus even though the dust is totally evacuated, the left-over gas metallicity is not significantly lower than the original (gas and dust) metallicity.

In the dust evacuated gas, fragmentation would be suppressed because the dust cooling and ${\rm H_{2}}$ formation on grain surfaces are quenched.
Therefore, more massive stars tend to form in such regions.

It is theoretically expected that star-forming clumps can fragment to dense cores of sub-solar mass when the metallicity is higher than the critical value $Z_{\rm crit} = 10^{-6} - 10^{-5}~Z_{\odot}$ \citep{Omukai2005, Schneider2012}.
However, the observed metallicity distribution of metal-poor stars is better reproduced with the critical metallicity at $Z_{\rm crit} = 10^{-4}~Z_{\odot}$ \citep{Salvadori2007}, much higher than the theoretical expectation.
Reduction of depletion factor $f_{\rm dep}$ in low-metallicity environments may explain this difference in the theoretical and empirical critical metallicities $Z_{\rm crit}$. 
This may be caused by slow dust growth in low-metallicity environments \citep{Asano2013,Remy-Ruyer2014}. 
We expect that the dust evacuation also contributed to reduce the depletion factor.
At the metallicity $Z \sim Z_{\rm crit}$, the dust-to-gas ratio $\mathcal D$ is easily reduced by the dust evacuation below the critical value $\mathcal D_{\rm crit}$.
In order for $\mathcal D$ to exceed $\mathcal D_{\rm crit}$, the metallicity needs to be higher than the theoretical critical metallicity in the dust-evacuated gas.
Therefore, the dust evacuation boosts the critical metallicity $Z_{\rm crit}$ compared with the theoretical prediction for $f_{\rm dep}$ similar to the Galactic value.

As the dust size decreases, the radiation force on dust grains becomes weaker due to the lower absorption efficiency. 
Therefore, small dust grains with the size less than $\sim 0.01\mu {\rm m}$
remain within star-forming clouds despite the stellar radiation feedback. 
\citet{Nozawa2007} and \citet{Schneider2012} indicated that the small grains of $\lesssim 0.01 \rm \mu m$
could be destructed in the reverse shock in SN remnants.

Cosmic rays (CRs) and Magnetic fields can suppress the dust evacuation.
CRs raise the ionization degree higher than $\sim 10^{-5}$, if the source galaxy with SFR $\gtrsim 1 ~M_{\odot}{\rm yr^{-1}}$ is within $40~{\rm kpc}$ or SN remnants are inside the same galaxy. 
In this case, the dust evacuation is significantly suppressed due to the Coulomb drag.
Also, the magnetic fields can be stronger than $0.1 ~\mu {\rm G}$ via turbulent small-scale dynamo amplification \citep{Schober2012}, and suppress the dust drift in the vertical direction.

A large fraction of observed metal-poor stars are CEMP stars \citep{Frebel2015,Aoki2006,Aoki2013,Yoon2016}.
For formation of CEMP stars, such scenarios as accretion of gases with metals but devoid of dust onto metal-poor stars from interstellar medium \citep{Johnson2015}, mass transfer from the companion star in the binary system \citep{Komiya2007} or metal enrichment from the faint SNe of the primordial stars \citep{Nomoto2013} are proposed.
The dust evacuation can also explain the observed CEMP star composition by altering the chemical composition of the gas from the original value.
The degree of depletion into grains sensitively depends on the metal species: 
for example, iron is more depleted into grains than carbon, according to the observations of interstellar medium \citep[e.g.,][]{Jenkins2009,Johnson2015}. 
Therefore, the dust evacuation can enhance the carbon abundance relative to iron.

\section*{Acknowledgements}
The authors wish to express our cordial thanks to Profs Takahiro
Tanaka and Takashi Hosokawa for their continual interest and advices.
We also thank Daisuke Nakauchi, Kazu Sugimura and Ryoki Matsukoba for for fruitful discussions.
We would like to thank to an anonymous reviewer for the constructive comments and advices, especially 
for Sec. 4.
This work is supported in part by MEXT/JSPS KAKENHI grants (KO:25287040, 17H01102, HY:17H04827).




\begin{thebibliography}{99}
\bibitem[\protect\citeauthoryear{Aoki et al.}{2006}]{Aoki2006}
Aoki W., et al., 2006, ApJ, 639, 897
\bibitem[\protect\citeauthoryear{Aoki et al.}{2013}]{Aoki2013}
Aoki W., et al., 2013, ApJ, 145, 13 
\bibitem[\protect\citeauthoryear{Asano et al.}{2013}]{Asano2013}
Asano R. S., Takeuchi T. T., Hirashita H., Nozawa T., 2013, MNRAS, 432, 637
\bibitem[\protect\citeauthoryear{Akimkin et al.}{2015}]{Akimkin2015}
Akimkin V. V., Kirsanova M. S., Pavlyuchenkov Ya. N., Wiebe D. S., 2015, MNRAS, 449, 440
\bibitem[\protect\citeauthoryear{Akimkin et al.}{2017}]{Akimkin2017}	
Akimkin V. V., Kirsanova M. S., Pavlyuchenkov Ya. N., Wiebe D. S., 2017, MNRAS, 469, 630
\bibitem[\protect\citeauthoryear{Barkana \& Loeb}{2001}]{Barkana2001}
Barkana R., Loeb A., 2001, Phys. Rep., 349, 125	
\bibitem[\protect\citeauthoryear{Bianchi \& Schneider}{2007}]{Bianchi2007}
Bianchi S., Schneider R., 2007, MNRAS, 378, 973
\bibitem[\protect\citeauthoryear{Bl\"umer et al.}{2009}]{Blumer2009}
Bl\"umer J., Engel R., H\"orandel J. R., 2009, Prog. Part. Nucl. Phys., 63, 293
\bibitem[\protect\citeauthoryear{Bromm et al.}{2001}]{Bromm2001}	
Bromm V., Ferrara A., Coppi P. S., Larson R. B., 2001, MNRAS, 328, 969
\bibitem[\protect\citeauthoryear{Bromm \& Loeb}{2003}]{Bromm2003}	
Bromm V., Loeb, A. 2003, Nature, 425, 812
\bibitem[\protect\citeauthoryear{Chabrier}{2003}]{Chabrier2003}	
Chabrier G., 2003, ApJ, 586, 133,
\bibitem[\protect\citeauthoryear{Chiao \& Wickramasinghe}{1972}]{Chiao1972}	
Chiao R. Y., Wickramasinghe N. C., 1972, MNRAS, 159, 361
\bibitem[\protect\citeauthoryear{Chen et al.}{2015}]{Chen2015}	
Chen Y., Bressan A., Girardi L., Marigo P., Kong X., Lanza A., 2015, MNRAS, 452, 1068
\bibitem[\protect\citeauthoryear{Clark et al.}{2011}]{Clark2011}	
Clark P. C., Glover S. C. O., Klessen R. S., Bromm V., 2011, ApJ, 727, 110 
\bibitem[\protect\citeauthoryear{Crutcher}{1999}]{Crutcher1991}	
Crutcher R. M., 1999, ApJ, 520, 706
\bibitem[\protect\citeauthoryear{Draine}{2003}]{Draine2003}
Draine B. T., 2003, ApJ, 598, 1026
\bibitem[\protect\citeauthoryear{Draine}{2011a}]{Draine2011}
Draine B. T., 2011a, ApJ, 732, 100
\bibitem[\protect\citeauthoryear{Draine}{2011b}]{Draine2011a}
Draine B. T., 2011b, Physics of the Interstellar and Intergalactic
Medium. Princeton University Press: Princeton, NJ
\bibitem[\protect\citeauthoryear{Draine \& Lee}{1984}]{Draine1984}
Draine B. T., Lee H. M., 1984, ApJ, 285, 89
\bibitem[\protect\citeauthoryear{Draine \& Salpeter}{1979}]{DS1979}
Draine B. T., Salpeter E. E., 1979, ApJ, 231, 77
\bibitem[\protect\citeauthoryear{Draine \& Sutin}{1987}]{DS1987}
Draine B. T., Sutin B., 1987, ApJ, 320, 803
\bibitem[\protect\citeauthoryear{Dwek et al.}{2007}]{Dwek2007}
Dwek E., Galliano F., Jones A. P., ApJ, 662, 927
\bibitem[\protect\citeauthoryear{Eisenstein \& Hu}{1998}]{Eisenstein1998}
Eisenstein D. J., Hu W., 1998, ApJ, 496, 605
\bibitem[\protect\citeauthoryear{Eisenstein \& Hu}{1999}]{Eisenstein1999}
Eisenstein D. J., Hu W., 1999, ApJ, 511, 5
\bibitem[\protect\citeauthoryear{Ferrara et al.}{1991}]{Ferra1991}
Ferrara A., Ferrini F., Barsella B., Franco J., 1991, ApJ, 381, 137
\bibitem[\protect\citeauthoryear{Frebel \& Norris}{2015}]{Frebel2015}
Frebel A., Norris J. E., 2015, ARA\&A, 53, 631
\bibitem[\protect\citeauthoryear{Fukushima et al.}{2018}]{Fukushima2017}
Fukushima H., Omukai K., Hosokawa T., 2018, MNRAS, 473, 4754
\bibitem[\protect\citeauthoryear{Glover \& Brand}{2003}]{Glover2003}
Glover S. C. O., Brand P. W. J. L., 2003, MNRAS, 340, 210
\bibitem[\protect\citeauthoryear{Jenkins}{2009}]{Jenkins2009}
Jenkins E. B., 2009, ApJ, 700, 1299
\bibitem[\protect\citeauthoryear{Johnson}{2015}]{Johnson2015}	
Johnson J. L., 2015, MNRAS, 453, 2771
\bibitem[\protect\citeauthoryear{Hirano et al.}{2014}]{Hirano2014}
Hirano S., Hosokawa T., Yoshida N., Umeda H., Omukai K., Chiaki G., Yorke H. W., 2014, ApJ, 781, 60
\bibitem[\protect\citeauthoryear{Hirano et al.}{2015}]{Hirano2015}
Hirano S., Hosokawa T., Yoshida N., Omukai K., Yorke H. W., 2015, MNRAS, 448, 568
\bibitem[\protect\citeauthoryear{Hosokawa et al.}{2011}]{Hosokawa2011}
Hosokawa T., Omukai K., Yoshida N., Yorke H. W., 2011, Science, 334,1250
\bibitem[\protect\citeauthoryear{Hosokawa et al.}{2016}]{Hosokawa2016}
Hosokawa T., Hirano S., Kuiper R., Yorke H. W., Omukai K., Yoshida N., 2016, ApJ, 824, 119
\bibitem[\protect\citeauthoryear{Hummer \& Storey}{1987}]{Hummer1987}
Hummer D. G., Storey P. J., 1987, MNRAS, 224, 801
\bibitem[\protect\citeauthoryear{Inayoshi \& Omukai}{2011}]{Inayoshi2011}
Inayoshi K., Omukai K., 2011, MNRAS, 416, 2748
\bibitem[\protect\citeauthoryear{Ishiki et al.}{2018}]{Ishiki2018}
Ishiki S., Okamoto T., Inoue A. K., 2018, MNRAS, 474, 1935
\bibitem[\protect\citeauthoryear{Katz}{1992}]{Katz1992}
Katz N., 1992, ApJ, 391, 502
\bibitem[\protect\citeauthoryear{Koh \& Wise}{2016}]{Koh2016}
Koh D., Wise J. H., 2016, MNRAS, 462, 81
\bibitem[\protect\citeauthoryear{Komiya et al.}{2007}]{Komiya2007}
Komiya Y., Suda T., Minaguchi H., Shigeyama T., Aoki W., Fujimoto M. Y., 2007, ApJ, 658, 367
\bibitem[\protect\citeauthoryear{Kroupa}{2001}]{Kroupa2001}
Kroupa P., 2001, MNRAS, 322, 231
\bibitem[\protect\citeauthoryear{Larson}{1998}]{Larson1998}
Larson R. B., 1998, MNRAS, 301, 569
\bibitem[\protect\citeauthoryear{Machida et al.}{2008}]{Machida2008}
Machida M. N., Omukai K., Matsumoto T., Inutsuka S.,2008, ApJ, 677, 813
\bibitem[\protect\citeauthoryear{Mathis et al.}{1977}]{Mathis1977}
Mathis J. S., Rumpl W., Nordsieck K. H., 1977, ApJ, 217, 425
\bibitem[\protect\citeauthoryear{McKee}{1989}]{McKee1989}
McKee C. F., 1989, ApJ, 345, 782
\bibitem[\protect\citeauthoryear{McKee \& Ostriker}{1977}]{McKee1977}
McKee C. F., Ostriker J. P., 1977, ApJ, 218, 148
\bibitem[\protect\citeauthoryear{Mo et al.}{1998}]{Mo1998}
Mo H. J., Mao S., White S. D. M., 1998, MNRAS, 295, 319 
\bibitem[\protect\citeauthoryear{Morgan \& Edmunds}{2003}]{Morgan2003}
Morgan H. L., Edmunds M. G., 2003, MNRAS, 343, 427
\bibitem[\protect\citeauthoryear{Nomoto et al.}{2013}]{Nomoto2013}
Nomoto K., Kobayashi C., Tominaga N., 2013, ARA\&A, 51, 457
\bibitem[\protect\citeauthoryear{Nozawa et al.}{2007}]{Nozawa2007}
Nozawa T., Kozasa T., Habe A., Dwek E., Umeda H., Tominaga N., Maeda K., Nomoto K., 2007, ApJ, 666, 955
\bibitem[\protect\citeauthoryear{Omukai}{2000}]{Omukai2000}
Omukai K., 2000, ApJ, 534, 809
\bibitem[\protect\citeauthoryear{Omukai et al.}{2005}]{Omukai2005}
Omukai K., Tsuribe T., Schneider R., Ferrara A., 2005, ApJ, 626, 627
\bibitem[\protect\citeauthoryear{Oh \& Haiman}{2002}]{Oh2002}
Oh S. P., Haiman Z., 2002, ApJ, 569, 558
\bibitem[\protect\citeauthoryear{Osterbrock \& Ferland}{2006}]{Osterbrock2006}
Osterbrock D. E., Ferland G. J. 2006, in Astrophysics of Gaseous Nebulae and Active Galactic Nuclei, ed. D. E. Osterbrock, \& G. J. Ferland (2nd ed.;
Sansolito, CA: Univ. Sci. Books)
\bibitem[\protect\citeauthoryear{Planck Collaboration}{2016}]{Planck2016}
Planck Collaboration XIII., A\&A, 2016, 594, 13
\bibitem[\protect\citeauthoryear{Plume et al.}{1997}]{Plume1997}
Plume R., Jaffe D. T., Evans N. J. II., Mart\'in-Pintado J., G\'omez-Gonz\'alez J., 1997, ApJ, 476, 730
\bibitem[\protect\citeauthoryear{Pollack et al.}{1994}]{Pollack1994}
Pollack J. B., Hollenbach D., Beckwith S., Simonelli D. P., Roush T., Fong W., 1994, ApJ, 421, 615
\bibitem[\protect\citeauthoryear{R\'emy-Ruyer et al.}{2014}]{Remy-Ruyer2014}
R\'emy-Ruyer A., Madden S. C., Galliano F., et al., 2014, A\&A, 563, 31
\bibitem[\protect\citeauthoryear{Salvadori et al.}{2007}]{Salvadori2007}
Salvadori S., Schneider R., Ferrara A., 2007, MNRAS, 381, 647
\bibitem[\protect\citeauthoryear{Santoro \& Shull}{2006}]{Santoro2006}
Santoro F., Shull J. M., 2006, ApJ, 643, 26
\bibitem[\protect\citeauthoryear{Schaerer}{2002}]{Schaerer2002}
Schaerer D., 2002, A\&A, 382, 28
\bibitem[\protect\citeauthoryear{Schneider et al.}{2002}]{Schneider2002}
Schneider R., Ferrara A., Natarajan P., Omukai K., 2002, ApJ, 571, 30
\bibitem[\protect\citeauthoryear{Schneider et al.}{2003}]{Schneider2003}
Schneider R., Ferrara A., Salvaterra R., Omukai K., Bromm V., 2003, Nature, 422, 869
\bibitem[\protect\citeauthoryear{Schneider et al.}{2012}]{Schneider2012}
Schneider R., Omukai K., Bianchi S., Valiante R., 2012, MNRAS, 419, 1566
\bibitem[\protect\citeauthoryear{Schober et al.}{2012}]{Schober2012}
Schober J., Schleicher D., Federrath C., Glover S., Klessen R. S.,  Banerjee R., 2012, ApJ, 754, 99
\bibitem[\protect\citeauthoryear{Solomon et al.}{1987}]{Solomon1987}
Solomon P. M., Rivolo A. R., Barrett J., Yahil A., 1987, ApJ, 319, 730
\bibitem[\protect\citeauthoryear{Spitzer}{1978}]{Spitzer1978}
Spitzer L., 1978, Physical Processes in the Interstellar Medium. Wiley, New York 
\bibitem[\protect\citeauthoryear{Stacy \& Bromm}{2007}]{Stacy2007}
Stacy A., Bromm V., 2007, MNRAS, 382, 229
\bibitem[\protect\citeauthoryear{Stacy et al.}{2016}]{Stacy2016}
Stacy A., Bromm V., Lee A. T., 2016, MNRAS, 462, 1307
\bibitem[\protect\citeauthoryear{Susa et al.}{2014}]{Susa2014}
Susa H., Hasegawa K., Tominaga N., 2014, ApJ, 792, 32
\bibitem[\protect\citeauthoryear{Todini \& Ferrara}{2001}]{Todini2001}
Todini P., Ferrara A., 2001, MNRAS, 325, 726
\bibitem[\protect\citeauthoryear{Wolfire et al.}{1995}]{Wolfire1995}
Wolfire M. G., Hollenbach D., McKee C. F., Tielens A. G. G. M., Bakes E. L. O., 1995, ApJ, 443, 152
\bibitem[\protect\citeauthoryear{Weingartner \& Draine}{2001}]{WD2001}
Weingartner J. C., Draine B. T., 2001, ApJS, 134, 263
\bibitem[\protect\citeauthoryear{Yoon et al.}{2016}]{Yoon2016}
Yoon J., Beers T. C., Placco V. M., Rasmussen K. C., Carollo D., He S., Hansen T. T., Roederer I. U., Zeanah J., 2016, ApJ, 833, 20   
\end{thebibliography}




\appendix
\section{Effects of Coulomb drag}\label{sec.effect_of_Coulomb}\label{apd1}

In this paper, we only take the collisional drag force ($F_{\rm collision}$) into account in calculating the terminal velocity of dust grains. 
However, as the ionization fraction increases, the acceleration of dust grains will be regulated by the Coulomb drag force ($F_{\rm Coulomb}$). 
 Here we investigate the effect of the Coulomb drag on the terminal velocity of dust grains. 
 The total drag force is given by \citep{DS1979}:
\begin{eqnarray}
	F_{\rm drag} &=& F_{\rm collision} + F_{\rm Coulomb} \nonumber \\
	                     &=& 2 \pi a^2 k T n_{\rm H} \left[ G_{0}(s) + x_{\rm e} \phi^2 \ln ( \Lambda) G_{2} (s) \right]  ,\label{0626.1} 
\end{eqnarray}
where
\begin{eqnarray}	                
	G_{0} &=& \frac{8 s}{3 \sqrt{\pi}} \left( 1 + \frac{9 \pi}{64} s^2 \right)^{1/2},  \label{0626.2} \\
	G_{2} &=& \frac{s}{(3 \sqrt{\pi}/4 + s^3 )}, \label{0626.3} \\
	\phi    &=& e U / k T, \label{0626.4} \\
	s        &=& (m v_{d}^2/2kT)^{1/2}, \label{0626.5} \\
	 \Lambda &=& \frac{3}{2 a e |\phi|} \left( \frac{k T}{\pi x_{\rm e}n_{\rm H}}  \right)^{1/2}, \label{0626.6}
\end{eqnarray}
$x_{\rm e}$ is the ionization fraction and $U = z_{\rm grain} e /a$ is Coulomb potential of dust grains. 
The Coulomb potential is $U = z_{\rm grain} e /a$, where $z_{\rm grain}$ is the dust charge. 
Here we set the gas temperature at $T = 100 ~{\rm K}$. 
Below, we evaluate the terminal velocities of grains for different values of ionization fraction. 

The Coulomb drag sensitively depends on the grain charge,
which is determined by the balance between the following three processes;
(1) ion collisions $J_{\rm ion}$, (2) electron collisions $J_{\rm e}$,
(3) photoelectric effect $J_{\rm pe}$:
\begin{eqnarray}
	J_{\rm pe} (z_{\rm grain}) + J_{\rm ion} (z_{\rm grain}) = J_{\rm e} (z_{\rm grain}). \label{0706.2}
\end{eqnarray}
The rate of collisional effect is given as \citep{WD2001}  
\begin{eqnarray}
	J_{\rm i} = n_{\rm i} f_{\rm i} \sqrt{ \frac{8 k T}{\pi m_{\rm i}} } \pi a^2 \tilde{J} \left( a, T,  z_{\rm grain}\right), \label{0707.1}
\end{eqnarray}
where $f_{\rm i}$ is the probability that gas particles deliver charge to dust grains when they collide.
We use $f_{\rm i}=0.5$ for electrons and $1$ for ions \citep{Akimkin2015}.
Expression for $\tilde{J}$ in Equation \eqref{0707.1} is given by \citep{DS1987}:
\begin{eqnarray}
	\tilde{J} (\tau, \nu)  = \begin{cases}
 	 1 + \left( \frac{\pi}{2 \tau} \right) ^{1/2}  & \text{ $( \nu = 0 )$ } \\
	 \left[ 1 - \frac{\nu}{\tau} \right] \left[ 1 + \left( \frac{2}{\tau - 2 \nu} \right)^{1/2} \right]  & \text{  $( \nu < 0 )$} \\
	 \left[ 1 + (4 \tau + 3 \nu)^{-1/2}  \right]^2 \exp(- \theta_{\nu} / \tau)  & \text{ $( \nu > 0 )$}
	 \end{cases} \label{0707.4} 
\end{eqnarray}
where $\tau$, $\nu$ and $\theta_{\nu}$ are
\begin{eqnarray}
	\tau & = & \frac{a k T }{q_{\rm i}^2}, \label{0707.5} \\
	\nu  & = & \frac{Z e}{q_{\rm i}}, \label{0707.6} 
\end{eqnarray}
and
\begin{eqnarray}
	\theta_{\nu} & = & \frac{\nu}{1 + \nu^{-1/2}}. \label{0707.7}
\end{eqnarray}
In Equation \eqref{0707.5} and \eqref{0707.6}, $q_{\rm i}$ represents the charge of dust grains, and we use $q_{\rm i} =1$ here.
Also, the photoelectric rate is given by
\begin{eqnarray}
	J_{\rm pe} = \pi a^2 \int d \nu \left( \frac{c u_{\nu}}{h \nu} \right) Y_{\rm pe}(\nu, a, Z) Q_{\rm abs}(\nu),  \label{0707.8}
\end{eqnarray}
where $Y_{\rm pe}$ is photoelectric yield \citep{WD2001}, 
$u_{\nu}$ is monochromatic radiation energy density.
As in Figure \ref{zu0807.6}, the efficiency factor $Q_{\rm abs}$ for absorption is calculated by Mie theory using the dielectric function given by \citet{Draine1984} and \citet{Draine2003}.

\begin{figure}
    \begin{center}
      \includegraphics[width=\columnwidth]{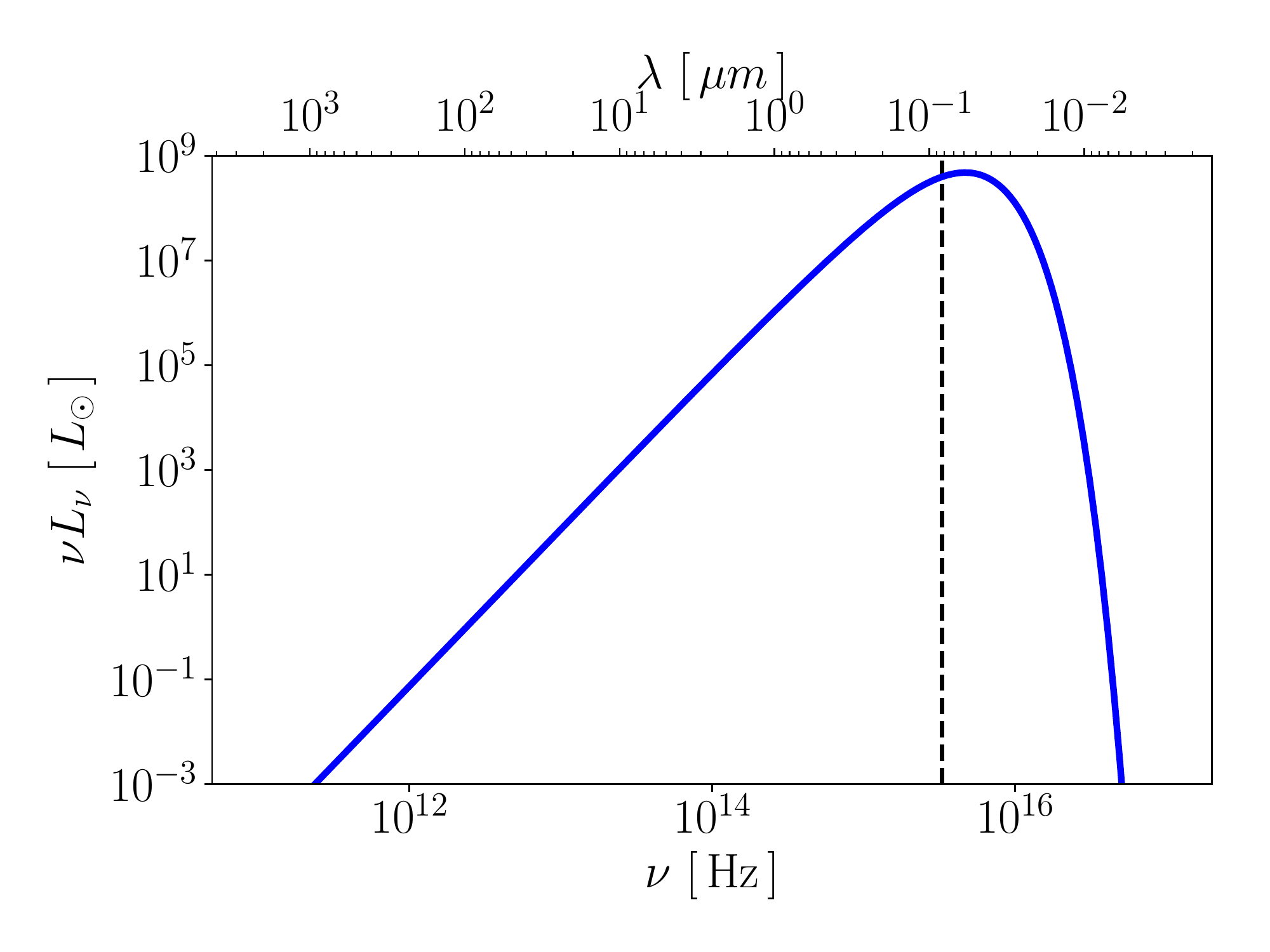}
        \caption{ 
        Spectral energy distribution of a star cluster formed in a star-forming cloud with mass $M_{\rm cl} = 10^{6}~M_{\odot}$, radius $R_{\rm cl} = 10 ~{\rm pc}$, and SFE $\epsilon_{*} = 0.1$. We set $m_{\rm ch} =10 ~M_{\odot}$ for Larson IMF.
        The vertical dashed line represents Lyman limit $0.0912~\mu {\rm m}$.
         \label{zu0702.2}}
         \end{center}
\end{figure}

We estimate $u_{\rm \nu}$ by summing up the contributions from stars according to the IMF (Eq. \ref{3.1.1}).
Figure \ref{zu0702.2} shows the specific luminosity in the case with the SFE $\epsilon_{*} = 0.1$. 
With the specific luminosity, the radiation energy density is given as $u_{\nu} = L_{\nu} / (4 \pi r^2 c)$. 
Thus,
\begin{eqnarray}
	J_{\rm pe} 
        & = & \frac{\pi a^2 L_{*}}{4 \pi r^2} \left \langle \frac{Y_{\rm pe} Q_{\rm abs}}{h \nu} \right \rangle,    \label{0707.9}
\end{eqnarray}
where $L_{*}$ is the bolometric luminosity. 
The bracket represents the mean value in the frequency range $\nu_{\rm min} - \nu_{\rm max}$:
\begin{eqnarray}
	\left \langle \frac{Y_{\rm pe} Q_{\rm abs}}{h \nu} \right \rangle = \frac{ \int ^{\nu_{\rm max}}_{\nu_{\rm min}}  d \nu \left( \frac{L_{\nu}}{h \nu}\right) Y_{\rm pe}(\nu, a, Z) Q_{\rm abs}(\nu)  } {\int ^{\nu_{\rm max}}_{\nu_{\rm min}} d \nu L_{ \nu } } .
\end{eqnarray}
Here we take the Lyman-limit frequency as $\nu_{\rm max}$, assuming that Lyman-continuum photons cannot escape from the H{\sc ii} regions.
\begin{figure}
    \begin{center}
      \includegraphics[width=\columnwidth]{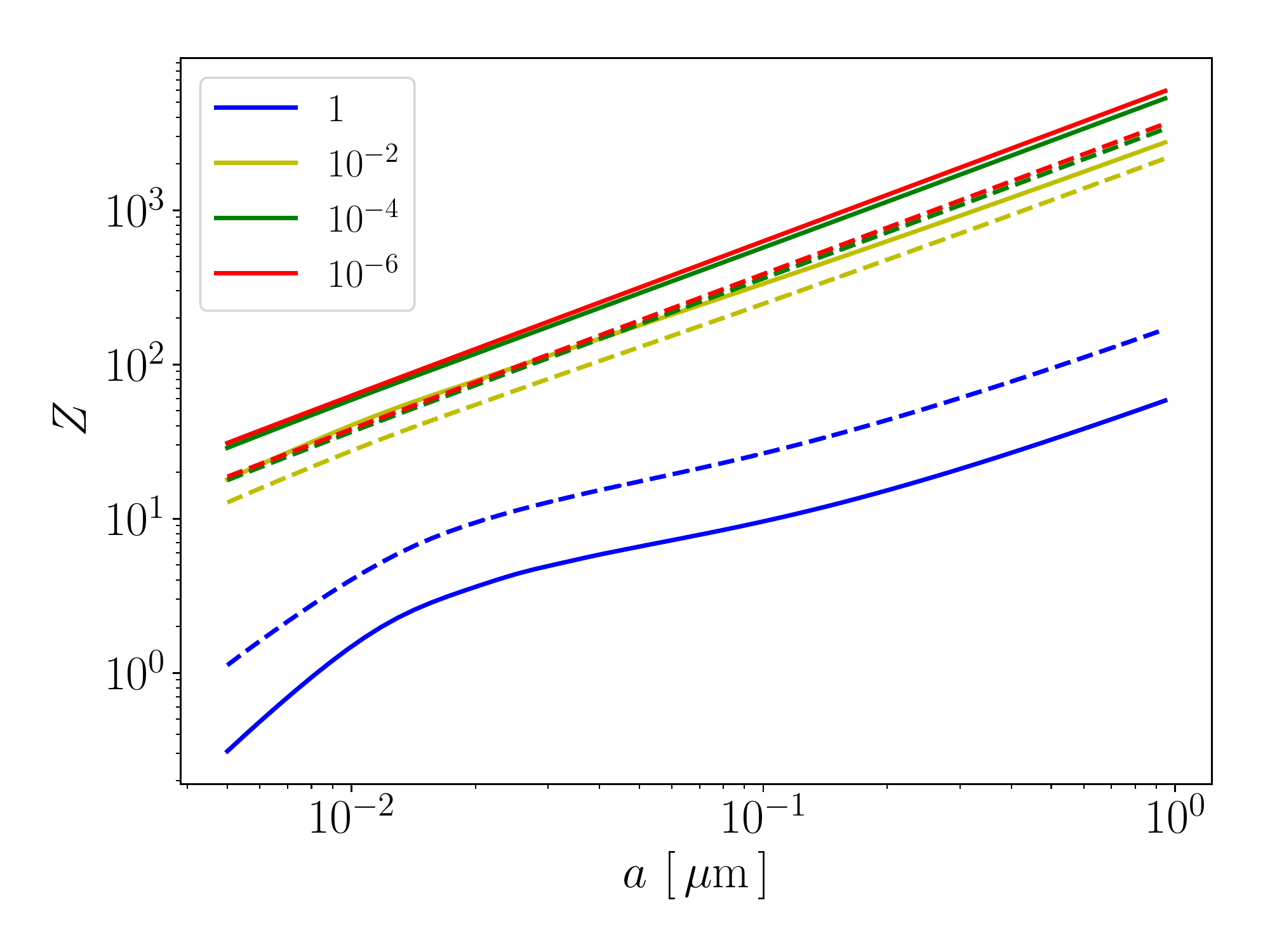}
        \caption{ 
         Grain charge as a function of grain radius. 
        The different lines correspond to the different ionization fractions, $x_{\rm e}  = 1$ (blue), $10^{-2}$ (green), $10^{-4}$ (yellow), and $10^{-6}$ (red).
         \label{zu.0708.1}}
         \end{center}
\end{figure}

In the following, we consider the cases of star-forming clouds with mass $M_{\rm cl} = 10^6~ M_{\odot}$ and radius $R_{\rm cl} = 10 ~\rm{pc}$, and set $\epsilon_{*}=  0.1$ and characteristic stellar mass $m_{\rm ch} = 10~M_{\odot}$ for the Larson IMF.
Dust grain charges are computed for four different ionization fractions $x_{\rm e} = 1, 10^{-2},10^{-4},$ and $ 10^{-6}$.
Figure \ref{zu.0708.1} shows the grain charge as a function of its size. 
The grain charge decreases as the ionization fraction increases. 
This is because free electrons are adsorbed to grains efficiently for the higher ionization fraction, thereby offsetting their photoelectrically induced positive charge. 
Yet, the energy of photons must exceed the Coulomb potential for grains to eject electrons.
The photoelectric process thus becomes inefficient for the Coulomb potential exceeding $\sim 13.6~\rm eV$. 
Even at $x_{\rm e} \sim 10^{-4}$, the Coulomb potential becomes close to the Lyman limit. 
Thus, the grain charge does not increase remarkably at $x_{\rm e} \lesssim 10^{-4}$.

\begin{figure}
    \begin{center}
      \includegraphics[width=\columnwidth]{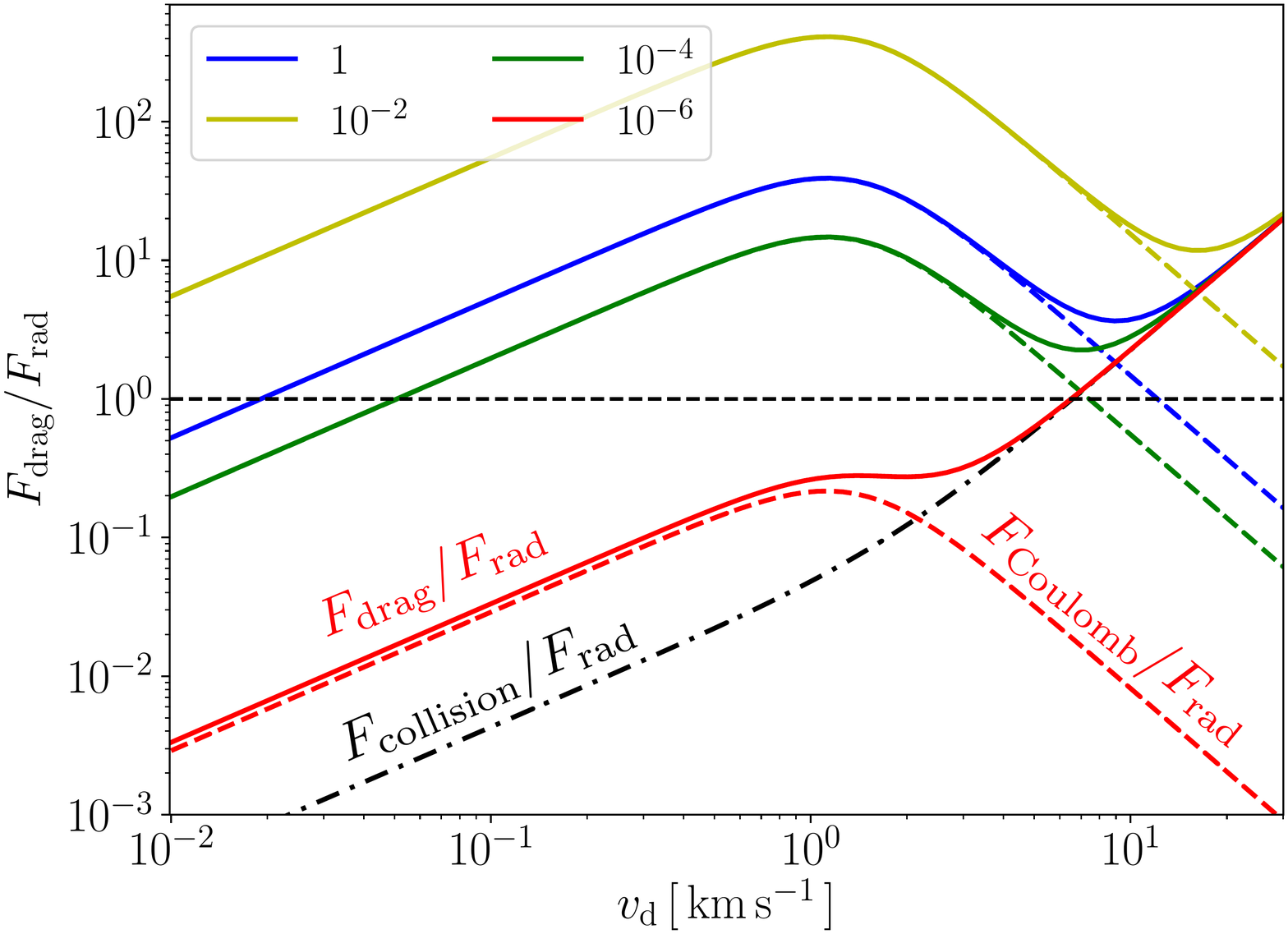}
        \caption{ 
        The ratio of the drag force to the radiation force on to a graphite grain of $0.1~{\rm \mu m}$.
        The different lines correspond to the different ionization fraction, $x_{\rm e} = 1$ (blue), $10^{-2}$ (green), $10^{-4}$(yellow), and $10^{-6}$ (red).
        The contribution by the Coulomb drag $F_{\rm Coulomb}$ and the collisional drag force $F_{\rm collision}$ are shown by dashed and dot-dashed lines, respectively.
        The horizontal dashed line represents the drag force being equal to the radiation force.
         \label{zu.1120}}
         \end{center}
\end{figure}

Figure \ref{zu.1120} shows the ratio of the drag force to the radiation force on to a graphite grain of $0.1~{\rm \mu m}$ as a function of the grain velocity.
With the ionization fraction higher than $10^{-4}$, the Coulomb drag becomes equal to the radiation force at $v_{\rm d} < 10^{-1}~{\rm km \, s^{-1}}$, 
and the grains cannot move fast enough to decouple from the gas component.
On the other hand, in the case of $x_{\rm e} = 10^{-6}$, the Coulomb drag is minor and the collisional drag instead balances with the radiation force.
In this case, we can calculate the terminal velocity of grains, ignoring the Coulomb drag force, as in Section  \ref{sec.det}.
\begin{figure}
    \begin{center}
      \includegraphics[width=\columnwidth]{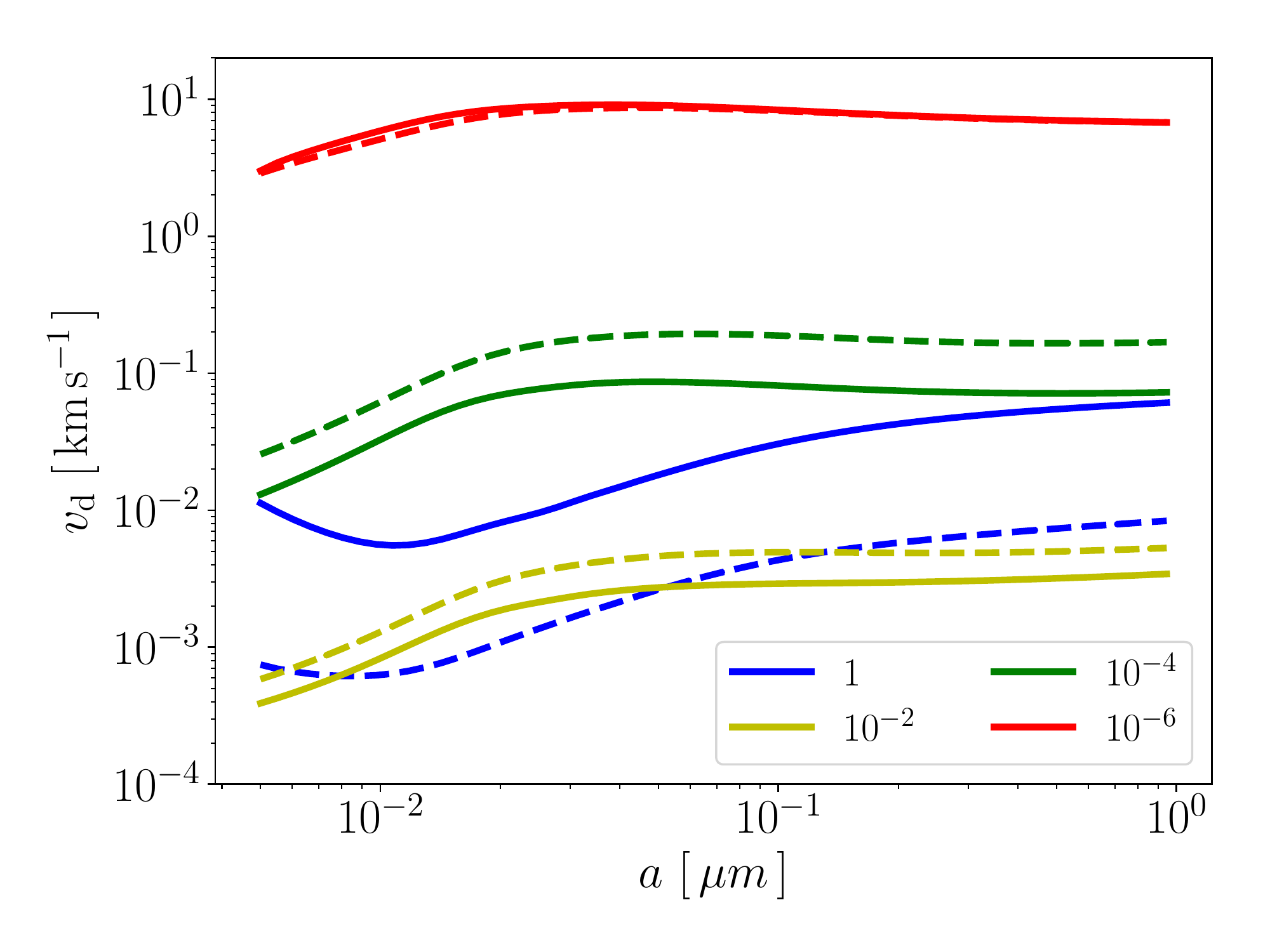}
        \caption{ 
        Terminal velocity of grains as a function of their size.
        	The solid and dashed lines represent graphite and silicate grains, respectively.
	The different lines correspond to the different ionization fractions, $x_{\rm e}  = 1$ (blue), $10^{-2}$ (green), $10^{-4}$ (yellow), and $10^{-6}$ (red).
	\label{zu.0730.2}}
         \end{center}
\end{figure}
\begin{figure}
 	\begin{center}
	\includegraphics[width=\columnwidth]{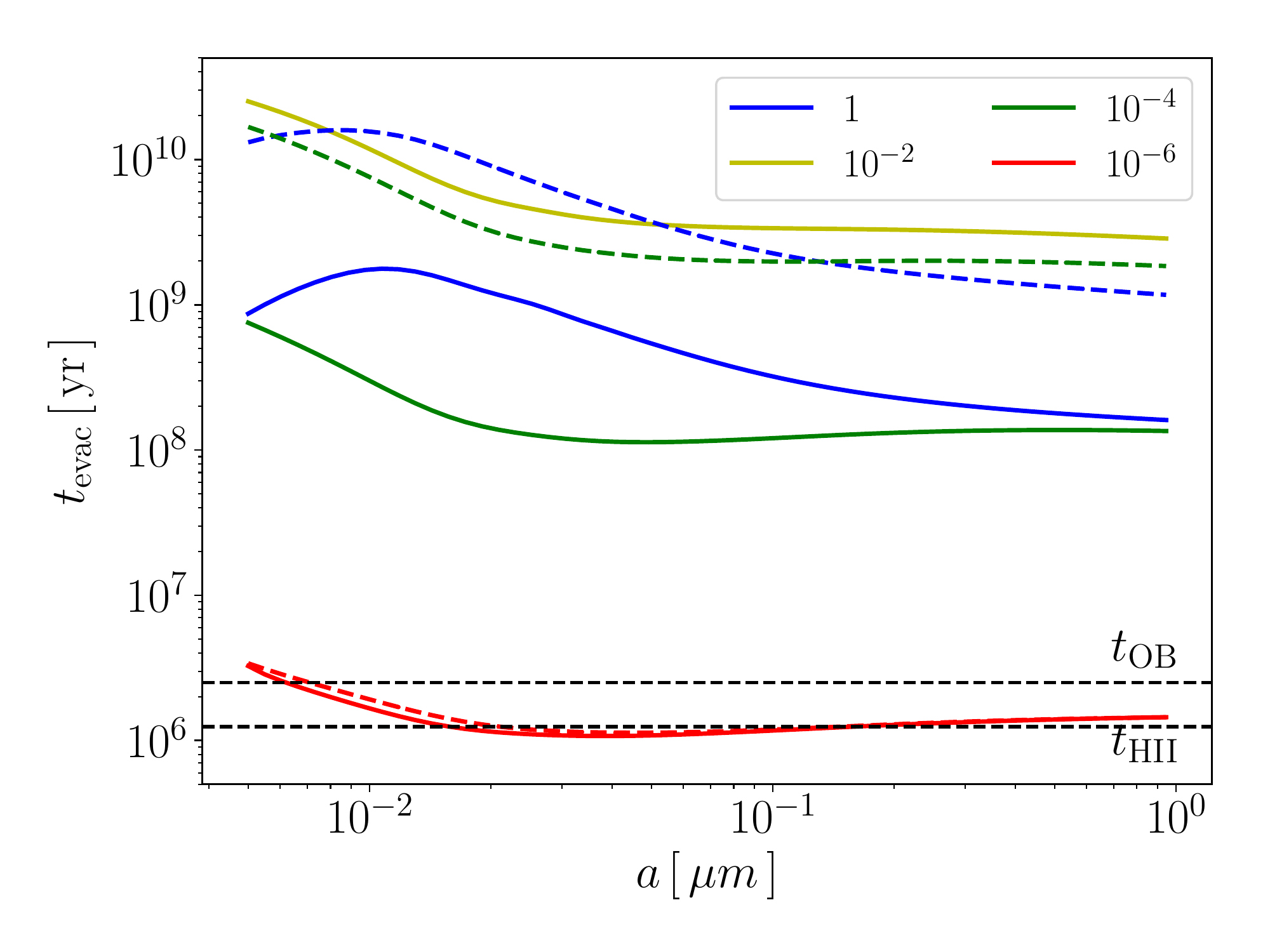}
	\caption{ 
	Dust evacuation time $t_{\rm evac}$ as a function of the grain size in the case with $(M_{\rm cl}, R_{\rm cl}, \epsilon_{*}) = (10^{6}~M_{\odot}, 10~ {\rm pc}, 0.1)$.
	As in Figure \ref{zu.0730.2}, colors correspond to different ionized fractions. 
	Also shown are the expansion time of H{\sc ii} regions $t_{\rm HII}$ and lifetime of OB stars $t_{\rm OB}$.
       \label{zu.0801.1}}
	\end{center}
\end{figure}
Figure \ref{zu.0730.2} shows the terminal velocity of grains, set by the balance $F_{\rm rad} = F_{\rm collision} + F_{\rm Coulomb}$, as a function of the size.
As expected, the terminal velocity is significantly reduced due to the Coulomb drag when $x_{\rm e} > 10^{-6}$.
At $a \lesssim 0.1~\rm \mu m$, the terminal velocity decreases for smaller dust size,
because of the lower radiation absorption efficiency (Fig. \ref{zu0807.6}).
Figure \ref{zu.0801.1} shows the dust evacuation time $t_{\rm evac}$ calculated by Equation \eqref{1.6.4}.
At $x_{\rm e} = 10^{-6}$, grains in the size range $2 \times 10^{-2} - 10^{-1}~\mu {\rm m}$ satisfy the condition of dust evacuation, 
while these of $a < 2\times 10^{-2} ~\mu {\rm m}$ ( or $a >10^{-1} ~\mu {\rm m}$) does not.
Therefore, small grains may remain in the star-forming region if the size distribution extends over a wide range.


\bsp	
\label{lastpage}
\end{document}